\documentclass[11pt, authoryear]{article}
\usepackage[colorlinks,citecolor=blue,urlcolor=blue]{hyperref}
\usepackage{wrapfig,lipsum,booktabs}
\usepackage{enumitem}
\usepackage{amsfonts, graphicx,color,multirow, amsthm,amsmath,natbib}
\usepackage{subcaption}
\usepackage{setspace,url}
\usepackage{multirow}
\usepackage{verbatim}
\usepackage[ruled,linesnumbered]{algorithm2e}
\def\boxit#1{\vbox{\hrule\hbox{\vrule\kern6pt\vbox{\kern6pt#1\kern6pt}\kern6pt\vrule}\hrule}}

\renewcommand{\baselinestretch}{1.45}

\newtheorem{theorem}{Theorem}
\newtheorem{corollary}{Corollary}

\newcommand{\bb}{\mbox{\bf b}}

\newcommand{\bA}{\mbox{\bf A}}
\newcommand{\ba}{\mbox{\bf a}}

\newcommand{\br}{\mbox{\bf r}}
\newcommand{\bB}{\mbox{\bf B}}
\newcommand{\bC}{\mbox{\bf C}}

\newcommand{\bH}{\mbox{\bf H}}
\newcommand{\bh}{\mbox{\bf h}}
\newcommand{\bI}{\mbox{\bf I}}
\newcommand{\bK}{\mbox{\bf K}}

\newcommand{\bM}{\mbox{\bf M}}

\newcommand{\bR}{\mbox{\bf R}}

\newcommand{\bV}{\mbox{\bf V}}

\newcommand{\bW}{\mbox{\bf W}}

\newcommand{\bZ}{\mbox{\bf Z}}
\newcommand{\bone}{\mbox{\bf 1}}

\newcommand{\bepsilon}{\mbox{\boldmath $\epsilon$}}

\newcommand{\bxi}{\mbox{\boldmath $\xi$}}

\newcommand{\bmu}{\mbox{\boldmath $\mu$}}
\newcommand{\bgamma}{\mbox{\boldmath $\gamma$}}

\newcommand{\bSigma}{\mbox{\boldmath $\Sigma$}}

\newcommand{\FDR}{\mbox{FDR}}
\newcommand{\FDP}{\mbox{FDP}}
\newcommand{\FNR}{\mbox{FNR}}
\newcommand{\FNP}{\mbox{FNP}}

\newcommand{\cov}{\mathrm{cov}}

\newcommand{\argmin}{\mathrm{argmin}}

\def\cov{\mbox{cov}}

\def\bB{\mbox{\bf B}}
\def\spacingset#1{\renewcommand{\baselinestretch}%
{#1}\small\normalsize} 
\renewcommand{\appendix}{
 \setcounter{section}{0}%
  \setcounter{subsection}{0}%
  \renewcommand\thesection{\Alph{section}}
  \setcounter{equation}{0}
  \renewcommand{\theequation}{S.\arabic{equation}}
  \setcounter{figure}{0}
  \renewcommand\thefigure{S\arabic{figure}}
  \setcounter{table}{0}
  \renewcommand\thetable{S\arabic{table}}  
  }
\begin{document}
\title{\bf Skilled Mutual Fund Selection: False Discovery Control under Dependence 
}
\date{}
\author{Lijia Wang, Xu Han and Xin Tong \thanks{Lijia Wang is a Ph.D. candidate in the Department of Mathematics at the University of Southern California, Los Angeles, CA; Xu Han is Assistant Professor in the Department of Statistical Science, Fox Business School, Temple University, Philadelphia, PA; Xin Tong is Assistant Professor in the Department of Data Sciences and Operations, Marshall Business School, University of Southern California, Los Angeles, CA. The corresponding author is Xu Han (Email: hanxu3@temple.edu).}}
\maketitle

\begin{abstract}
\begin{singlespace}
Selecting skilled mutual funds through the multiple testing framework has received increasing attention from finance researchers and statisticians. The intercept $\alpha$ of Carhart four-factor model is commonly used to measure the true performance of mutual funds, and positive $\alpha$'s are considered as skilled. We observe that the standardized OLS estimates of $\alpha$'s across the funds possess strong dependence and nonnormality structures, indicating that the conventional multiple testing methods are inadequate for selecting the skilled funds. We start from a decision theoretic perspective, and propose an optimal multiple testing procedure to minimize a combination of false discovery rate and false non-discovery rate. Our proposed testing procedure is constructed based on the probability of each fund not being skilled conditional on the information across all of the funds in our study. To model the distribution of the information used for the testing procedure, we consider a mixture model under dependence and propose a new method called ``approximate empirical Bayes" to fit the parameters. Empirical studies show that our selected skilled funds have superior long-term and short-term performance, e.g., our selection strongly outperforms the S\&P 500 index during the same period. 
\end{singlespace}
\end{abstract}
\vspace{1cm}
\vspace{1cm}

\section{Introduction}
The mutual fund industry is giant, with trillions of dollars of assets under management. It plays an important role in shaping the economy by pooling money from many investors (retail investors as well as institutional investors) to purchase securities. Meanwhile, the industry has a great impact on household finances. \citet{MXZ20} shows that mutual funds in the past decade have become a more important provider of liquidity to households. In 2019, 46.4\% of the households owned funds in the United States according to  Statista\footnote{The statistics are available at \url{https://www.statista.com}}. While the success of the fund industry is commonly attributed to several advantages: diversification (reduce risk), marketability (easy access), and professional management, the success of individual fund mainly depends on the fund manager's capability. 

How to select mutual funds with true capabilities to make profits is a challenging and substantial problem in finance. Those funds with good performance during the prior few years do not guarantee profits for the subsequent years. For example, \verb!Rydex Series Funds:! \verb!Inverse S&P 500 Strategy Fund! generated a total return of 383.3\% between 2001 and 2005, but it had a total return of -96.2\% between 2006 and 2019. Even in a less extreme case, considering the size of mutual funds, an unskilled mutual fund that loses a small proportion of its investment could incur a huge loss in terms of the dollar amount. 

\citet{Carhart97} four-factor model has been widely used to evaluate a mutual fund's performance. Regressing the excess return on four market factors, the intercept $\alpha$ in the cross-sectional linear regression model is considered as a measure of the fund's true capability of making profits. By convention, $\alpha>0$ is classified as \textit{skilled}, $\alpha<0$ as \textit{unskilled} and $\alpha=0$ as \textit{zero-alpha}. In practice, the true value of $\alpha$ is not observed. Instead, the ordinary least squares (OLS) estimator of $\alpha$ is calculated for the comparison between different funds. Some funds may have a zero or even a negative $\alpha$, meaning that they cannot make profits by adjusting the market factors.  Due to luck, however, their estimated alpha's can be positive for some years. Investment in such funds can lead to substantial loss rather than profits in the future. There is an urgent demand to develop strategies for selecting truly skilled mutual funds while avoiding or controlling the false selection of those ``lucky" funds (unskilled or zero-alpha funds). The seminal paper \citet{BSW10} initiated a multiple testing framework for the fund selection problem. Selecting the skilled mutual funds can be formulated as simultaneously testing $H_{i0}: \alpha_i\leq 0\quad\text{vs.}\quad H_{ia}: \alpha_i>0$ for thousands of funds based on the test statistics as the standardized OLS estimates of $\alpha$'s. By adopting the false discovery rate (FDR) concept from \citet{BH95}, \citet{BSW10} applied \citet{Storey02} procedure for selecting the skilled funds, aiming to control the FDR at a certain level,  where FDR is defined as the expected proportion of false selection of ``lucky" funds among the total selection. 

When we consider funds during a certain period, the test statistics are usually dependent. The dependence can be very strong and can greatly affect the FDR control. For example, such strong dependence has a huge impact on funds' locations in the empirical distribution of the test statistics, which can cause small p-values for ``lucky" funds (false selection), or relatively large p-values for skilled funds (mis-selection). Consequently, the selection of ``lucky" funds may increase while more skilled funds will be ignored. \citet{Storey02} procedure is usually valid for FDR control under weak dependence settings but not for strong dependence. Another issue in the aforementioned multiple testing is that p-values are usually calculated based on the assumption that each test statistic is normally distributed. In our fund data, we observe nonnormality and the strong dependence of the test statistics, indicating that the method in \citet{BSW10} is not ideal.

To address the above challenging issues arising from the mutual fund data, we consider the conditional probability of each fund not being skilled given the test statistics across all of the funds in our study. For convenience, we call such conditional probability as \textit{the degree of non-skillness} (d-value).  Our multiple testing procedures will be constructed based on the d-value, in contrast with the p-value based selection strategies.  Intuitively, d-values with smaller values suggest larger chances of being skilled and thus encouraging selection. The dependence and nonnormality structure of the test statistics are also preserved in the conditional information, so d-values can provide a good ranking for fund performance, which is crucial to fund selections by our multiple testing procedure. 

To construct the multiple testing procedure, we start from a decision theoretic perspective. In the decision process, in addition to the falsely selected lucky funds (false discoveries), another type of mistake is the mis-selected skilled funds (false non-discoveries). Control of the false discovery rate (FDR) has been a prevailing problem in large scale multiple testing (\cite{BH95}, \cite{BY01}, \cite{S02}, etc). Results on the control of the false non-discovery rate (FNR) are fewer yet equally important (\cite{S06}, \cite{SZG08}, etc). Consideration of one type of error alone can lead to unsatisfactory testing procedures. For example, a procedure with FDR control only may still be very conservative, with substantial power loss. A natural step is to minimize a combination of FDR and FNR by an optimal testing procedure. However, this problem has not been well studied in the statistics literature. Alternative attempts to this question can be found in \cite{GW02}, \cite{SC07} and subsequent papers in the direction where they considered marginal FDR and marginal FNR. When the test statistics possess general dependence, especially strong dependence, marginal FDR (FNR) is not equivalent to FDR (FNR). Our optimal testing procedure (Theorem~\ref{thm:opt.decision}) provides a solution to this long-standing problem by directly targeting the FDR and FNR. The d-values defined above are the key to achieving optimality.

To calculate the d-values, we need to model the empirical distribution of the test statistics. We consider a three-part mixture model to capture the dependence and nonnormality phenomenon in the test statistics. The mixture model is a powerful tool in high-dimensional statistics to model a multi-modal structure. In the existing literature, it is usually assumed that individual statistics of primary interest are independent; thus the joint likelihood can be decomposed as the product of marginal likelihoods to facilitate statistical analysis. However, such an independent assumption is not satisfied in our data. Another challenging issue in this mixture model is that there are seven free parameters. Conventional techniques such as empirical Bayes will require seven equations to solve those parameters. It is not desirable in our data analysis, since more equations will increase the difficulty of solving the parameters, and higher moments will also have more complicated expressions of the parameters. Instead, we propose a new method called ``approximate empirical Bayes" to fit the parameters here; it borrows the strength of empirical Bayes but also relieves the computational burden for more parameters. It shows that the mutual fund data in our study can be well modeled by our newly proposed method. 

The innovation of the above statistical procedures has shed new light on selecting skilled mutual funds. We first demonstrate the short-term out-of-sample performance of our selected skilled funds by a dynamic trading strategy. Hypothetically, if we invested \$1 million at the beginning of 2010 and then at the end of 2019, our portfolio would grow to \$3.52 million. In comparison, the portfolio by investing in the S\&P 500 index would grow to \$2.90 million, and the portfolios selected by other well-known multiple testing procedures \citep{BH95,Storey02}  would produce values less than \$1.75 million. By our testing procedure, based on the data for a certain period, we can classify funds into three groups: skilled funds, unskilled funds, and the rest as non-selected. We further evaluate the long-term performance of the three groups through subsequent moving 10-year windows. It shows that there are clear distinctions in the performance among the three groups over time and our selected skilled funds have better performance compared with the other groups. The superior performance of our selection attributes to the advantage of d-values in our procedure, in contrast with the p-value based strategies used in the above classical multiple testing procedures. To illustrate the difference between the two measurements of significance, we compare funds with the top 50 smallest p-values and those with the top 50 smallest d-values. Based on the estimated annual $\alpha$'s for the subsequent moving 10-year windows, the latter group has persistent long-term outperformance, while the performance of funds with the top 50 smallest p-values is extremely unstable. Such comparison indicates that the d-values provide a better ranking criterion for the funds' performance.

The mutual fund problem has received increasing attention from the finance researchers. The existing finance literature has also addressed some of the challenging issues in the mutual fund problem with quite different approaches from ours. For example, in the framework of multiple testing,   \citet{FC21} and \citet{HL20} have also focused on the Type II error, while \citet{LD19} and \citet{GLX21} have proposed procedures to weaken the dependence effect. On the other hand, in  the framework of estimation, \citet{HL18} has proposed EM algorithm for the mixture model. We will compare our method with these existing approaches both theoretically and numerically. The out-of-sample performance shows that our method dominates these approaches. 
 
The rest of the paper is organized as follows.  Section~\ref{sec:procedure} introduces our multiple testing procedure.   Section~\ref{sec:data} describes the mutual fund data and challenging issues. Section~\ref{sec:conditional} develops an approximate empirical Bayes method to fit the parameters and constructs the degree of non-skillness (d-value). Section~\ref{sec:real} analyzes the mutual fund data by using the proposed methods. All the technical derivations, additional figures, and tables are relegated to the Supplementary Materials. 

\section{Optimal Multiple Testing Procedure}~\label{sec:procedure} 
\noindent In this section, we will propose an optimal multiple testing procedure under dependence, which will be applied to the mutual fund selection problem in later sections. Suppose we have a $p$-dimensional vector $\bZ$ with unknown mean vector $\bmu=(\mu_1,\cdots,\mu_p)^\top$ and a known covariance matrix $\bSigma$. We want to simultaneously test 
\begin{equation*}
H_{i0}: \mu_i\leq 0\quad\text{vs.}\quad H_{ia}: \mu_i>0,\quad\quad i=1,\cdots,p
\end{equation*}
based on $\bZ$. Without loss of generality, we can assume that all diagonal elements of $\bSigma$ are 1, since we can standardize $\bZ$ otherwise. 

\begin{table}[h!!]
\begin{center}\caption{\footnotesize Classification of tested hypotheses}\label{ar}
  \begin{tabular}{cccc}
         \hline\hline
                       &Number    &Number    &\\
         Number of     &not rejected  &rejected  &Total\\
         \hline
         True Null     &$U$         &$V$      &$p_0$\\
         False Null    &$T$         &$S$      &$p_1$\\
         \hline
                       &$p-R$       &$R$      &$p$\\
         \hline
   \end{tabular}
\end{center}
\end{table}
Table \ref{ar} is the classification of multiple hypothesis testing, where $R$ denotes the total number of rejections (discoveries), $V$ denotes the total number of false rejections (false discoveries), and $T$ denotes the total number of false acceptance (false non-discoveries). Define the false discovery proportion (FDP) and the false nondiscovery proportion (FNP) as
\begin{equation*}
\FDP=\frac{V}{R} \quad \quad \mbox{and} \quad\quad \FNP=\frac{T}{p-R} 
\end{equation*}
respectively, to measure the mistakes that we make in the decision process. For notational convenience, we define 0/0=0 throughout the current paper. Correspondingly, we define the false discovery rate as $\FDR=E[\FDP]$ and the false non-discovery rate as $\FNR=E[\FNP]$. $\FDR$ and $\FNR$ have been regarded as the Type I error and the Type II error respectively for multiple testing procedures \citep{S06}. 

From a decision theoretic perspective, an optimal testing procedure can be constructed to minimize the objective function:
\begin{equation}\label{e1}
\FNR+\lambda \FDR, 
\end{equation}
where $\lambda$ is a tuning parameter to balance the two types of errors. (Choice of $\lambda$ is a nontrivial task in practice, but it is not relevant in our main discussion here.) Let's treat $\lambda$ as a generic variable. Suppose we consider a decision vector $\ba=(a_1,\cdots,a_p)$, where $a_i=1$ if we reject the $i$th null hypothesis and $a_i=0$ otherwise. Therefore, a false discovery can be expressed as $a_i \bI_{\mu_i\leq 0}$ where $\bI$ is an indicator function, and  a false non-discovery can be expressed as $(1-a_i)\bI_{\mu_i>0}$. The objective function can be written as
\begin{equation*}
E\Big[\frac{\sum_{i=1}^p(1-a_i)\bI_{\mu_i>0}}{p-\sum_{j=1}^pa_j}+\lambda\frac{\sum_{i=1}^pa_i\bI_{\mu_i\leq0}}{\sum_{j=1}^pa_j}\Big]. 
\end{equation*}
The above expectation is over the likelihood of $\bZ$ and the prior distribution of $\bmu$. It is equivalent to minimize the expectation of loss function conditional on $\bZ$, that is, minimizing
\begin{equation}\label{eq20}
L(\mathbf{a}) = \sum_{i=1}^p\Big[\frac{(1-a_i)P(\mu_i>0|\bZ)}{p-\sum_{j=1}^pa_j}+\lambda\frac{a_iP(\mu_i\leq0|\bZ)}{\sum_{j=1}^pa_j}\Big]. 
\end{equation}
Without any delicate analysis, minimizing the objective function would involve optimization over $2^p$ choices of $\ba$.
To the best of our knowledge, the minimizer to \eqref{eq20} has not been shown in the existing literature. We notice that the number of rejections $\sum_{j=1}^pa_j$ only have $p + 1$ possible values. It inspires us to decompose the problem into $p + 1$ optimization tasks, which shares a similar rationale as divide and conquer strategies.  Our optimization procedure is presented in the following theorem.

\begin{theorem}\label{thm:opt.decision}
Denote $\emph{DONS}_{(1)}, \cdots, \emph{DONS}_{(p)}$ as the non-decreasing order of $P(\mu_i\leq 0|\bZ)$, $i=1,\cdots,p$. Define the action vector $\mathbf{a}_0=(a_1^{(0)},\cdots,a_p^{(0)})=(0,\cdots,0)$. Given a value of $\lambda$, for each $j\in\{1,\cdots,p\}$, let the action vector $\mathbf{a}_j = (a_1^{(j)}\,, \cdots\,, a_p^{(j)})$ where
\begin{equation*}
a_i^{(j)} =\begin{cases}
1,&\mbox{if } P(\mu_i \leq 0 \mid \bZ) \leq \emph{DONS}_{(j)}\,;\\
0,&\mbox{if } P(\mu_i \leq 0 \mid \bZ) > \emph{DONS}_{(j)}\,.
\end{cases}
\end{equation*}
Calculate the corresponding loss function
\begin{equation} \label{eq:opt.loss}
L(\mathbf{a}_j)\equiv\sum_{i=1}^p\Big[\frac{(1-a_i^{(j)})P(\mu_i>0|\bZ)}{p-j}+\lambda\frac{a_i^{(j)}P(\mu_i\leq0|\bZ)}{j}\Big]
\end{equation}
for $j=0,\cdots,p$. The optimal procedure for (\ref{eq20}) is $\mathbf{a}_{opt}=\argmin_{\{\mathbf{a}_j :  0\leq j\leq p\}}L(\mathbf{a}_j)$.
\end{theorem}
The detailed proof is available in Supplementary Materials~B. For the ease of presentation, we consider $\bmu$ as the mean vector. Theorem~\ref{thm:opt.decision} is still valid for a more general vector of parameters of interest. Note that if we restrict the total number of rejections to be $k$, the second term in the loss function~\eqref{eq:opt.loss} will be \begin{equation*}
\lambda\frac{\sum_{i=1}^pa_i^{(j)}P(\mu_i\leq 0|\bZ)}{j}=\lambda\frac{\sum_{i=1}^k\mbox{DONS}_{(i)}}{k}\,.
\end{equation*}
The term $k^{-1}\sum_{i=1}^k\mbox{DONS}_{(i)}$ can be regarded as an FDR conditional on $\bZ$. This motivates a FDR control procedure: given a $\theta$ level, we select the largest $k$ such that 
\begin{equation}\label{eq3}
\frac{1}{k}\sum_{i=1}^k \mbox{DONS}_{(i)}\leq \theta. 
\end{equation}
Instead of minimizing the objective function~\eqref{eq20} that involves the tuning parameter $\lambda$, this procedure targets the FDR control directly. Details of procedure \eqref{eq3} are presented in Algorithm~\ref{alg:general}. The $\theta$ level reflects our tolerance for making mistakes of false rejections. We should point out that a procedure only focusing on FDR control may be very conservative, since it can reject fewer hypotheses to avoid false rejections. Ideally, we want to hunt for more true rejections in addition to the FDR control. Fortunately, our procedure achieves this goal since the false non-discovery rate is minimized, which is resulted from the monotonicity established in the next result.

\begin{algorithm}[htb!]
\caption{Simplified Optimal Procedure} \label{alg:general}
\SetKw{KwBy}{by}
\SetKwInOut{Input}{Input}\SetKwInOut{Output}{Output}
\SetAlgoLined

\Input{The vector of test statistics $\bZ = (Z_1, \cdots, Z_p)$; the significance level $\theta$. }

Compute the d-values $P(\mu_i\leq 0|\bZ)$ \tcp*{Details of d-values are in Section~\ref{sec:conditional}.}

$(\mbox{DONS}_{(1)}, \cdots, \mbox{DONS}_{(p)})\leftarrow$ sort $\{P(\mu_i\leq 0|\bZ)\}$ in non-decreasing order\,;

$j \leftarrow \max\{k \mid k^{-1}\sum_{i=1}^k\mbox{DONS}_{(i)} \leq \theta\}$\,;

$ (a_1, \cdots a_p) \leftarrow (0, \cdots, 0)$\,;

\For{$i = 1,\cdots,p$}{
\If{$P(\mu_i\leq 0|\bZ) \leq \emph{DONS}_{(j)}$}{$a_i \leftarrow 1$\,;}
}

\Output{ the decision $(a_1, \cdots a_p)$. }
\end{algorithm}

\begin{corollary}\label{prop:monoton}
$k^{-1}\sum_{i=1}^k\emph{DONS}_{(i)}$ is monotonically nondecreasing in $k$, and \\ $(p-k)^{-1}\sum_{i=k+1}^p\emph{DONS}_{(i)}$ is monotonically nondecreasing in $k$. 
\end{corollary}


The proof is in Supplementary Materials B.  According to the decision $\ba_j$ in Theorem~\ref{thm:opt.decision},  the FNR conditional on $\bZ$ can be expressed as
\begin{equation*}
\frac{\sum_{i = 1}^p(1-a_i)P(\mu_i>0|\bZ)}{p-j}
= \frac{\sum_{i = k + 1}^p(1 - \mbox{DONS}_{(i)})}{p-k}\\
= 1-\frac{\sum_{i=k+1}^p\mbox{DONS}_{(i)}}{p-k}\,,  \mbox{ when } j = k \,.
\end{equation*}
Therefore, the FNR is monotonically nonincreasing in $k$ by Corollary~\ref{prop:monoton}. Combining the two results in Corollary~\ref{prop:monoton}, the FNR based on testing procedure~\eqref{eq3} is minimum among all the testing procedures based on the test statistics $\bZ$ while the FDR is controlled at the level $\theta$. 

The testing procedure (\ref{eq3}) is a well-known step-up procedure in the statistics literature, see \cite{TZ07}, \cite{SC07}, \cite{SZG08}, etc. We should also point out that optimality depends on the definition of the objective goal. For example, \cite{TZ07} showed that maximizing $R$ subject to $E[V/R]\leq\theta$ also leads to the testing procedure (\ref{eq3}). However, the connection of (\ref{eq3}) to the optimal testing procedure in Theorem~\ref{thm:opt.decision} has not been revealed in the existing literature. 

\section{Mutual Fund Data and Challenges}\label{sec:data}

\subsection{Equity Fund Data Description}
\begin{figure}[!ht]
\begin{center}
\scalebox{0.3}{\includegraphics{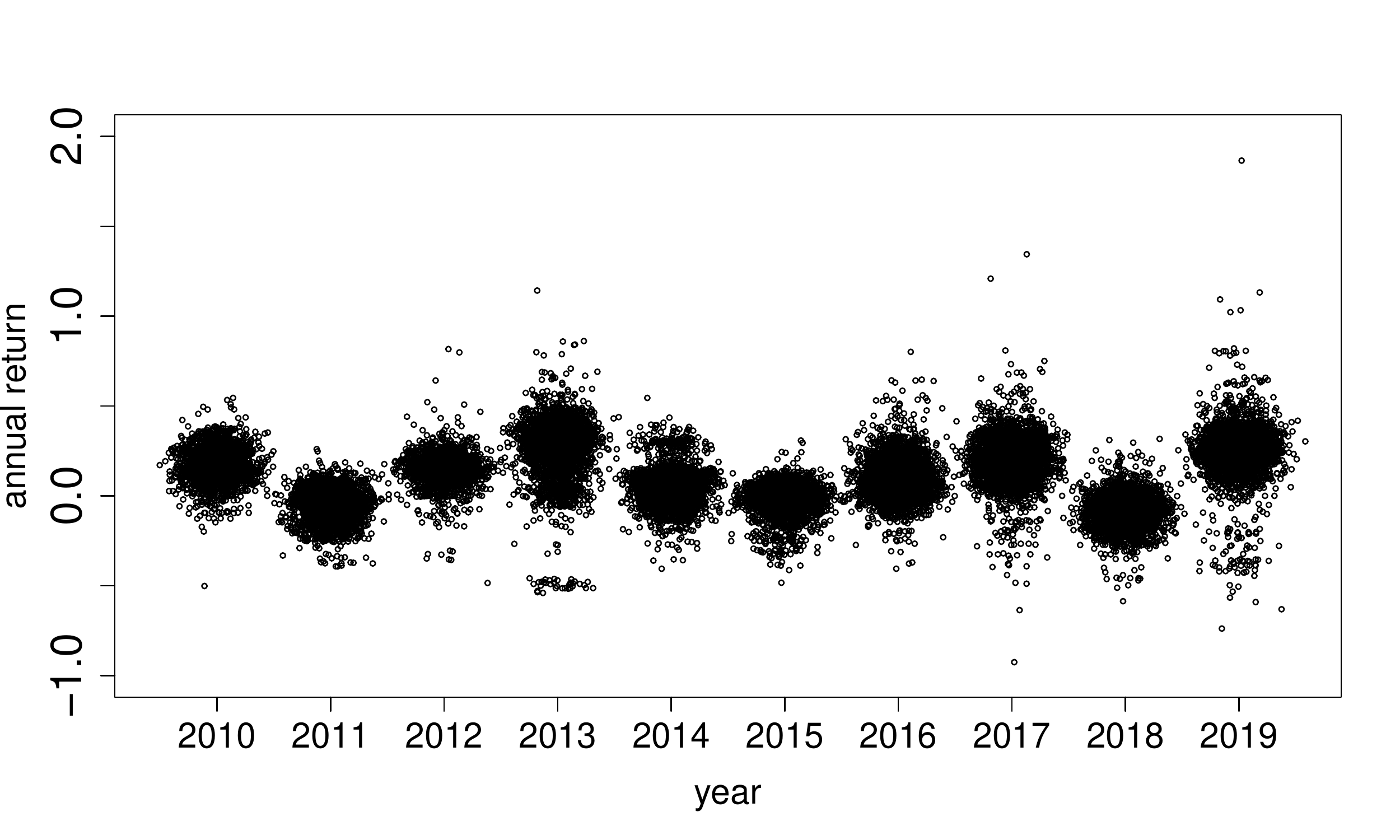}}
\end{center}
\vspace{-0.3cm}
\caption{\footnotesize Annual returns for the mutual funds existing in the 10-year windows from 2000-2009 to 2009-2018. Each cluster center at a certain year represents the annual returns in that year for equity funds that survived during the previous ten years. Moreover, the y-coordinate of each point in a certain cluster is the annual return in that year so that the number of funds with the same level of annual returns is observable. }\label{fig:annual.return}
\end{figure}

We download the monthly return data of equity mutual funds from the CRSP survivor-bias-free US mutual fund database available at the Wharton Research Data Services (WRDS)\footnote{The monthly return data are available at \url{https://wrds-web.wharton.upenn.edu/wrds/ds/crsp/mfund_q/monthly//index.cfm}}. The monthly returns are net of trading costs and expenses (including fees). We select the funds existing from the beginning of 2000 to the end of 2019 (20 years). Throughout the paper, we focus on the equity funds that survived for at least 10 years and analyze the equity funds with complete monthly return data during each 10-year periods (2000-2009, 2001-2010, $\cdots$, 2009-2018). We notice that WRDS uses ``0" to denote a missing value for the monthly return. To be cautious with missing values, we delete any funds with a monthly return as zero for each 10-year period. This data cleaning step reduces the number of mutual funds to several thousand for each 10-year period, e.g., $2{,}722$ equity funds during 2000-2009, and $5{,}123$ funds during 2009-2018. The reason to consider 10-year windows will be explained in Section~\ref{sec.sub:alpha}. We believe that the dataset with such a great amount of funds is sufficient for our analysis. There are three variables in this dataset: date, fund ID, and monthly return per share. For all the equity funds that survived in each 10-year period, Figure~\ref{fig:annual.return} plots their annual returns for the subsequent one year. For example, the cluster centered at 2010 represents the annual returns in 2010 for equity funds that survived during 2000-2009. It illustrates that the distributions of fund returns vary over the years, but most funds have annual returns around 0. Our primary goal here is to  figure out how to select funds with persistent future performance based on the historical data.  

To apply \citet{Carhart97} four-factor model  in Section~\ref{sec.sub:alpha}, we download data for the Fama-French market factors during this period from WRDS\footnote{The Fama-French market factor data are available at \url{https://wrds-web.wharton.upenn.edu/wrds/ds/famafrench/factors_m.cfm?navId=203}}. There are six variables: (1) date, (2) monthly excess return on the market which is calculated as the value-weight return on all NYSE, AMEX and NASDAQ stocks (from CRSP) minus the one-month Treasury bill rate (from Ibbotson Associates), (3) the monthly average return on the three small portfolios minus the average return on the three big portfolios, (4) the monthly average return on the two value portfolio (the high BE (Book Equity)/ ME (Market Equity) ratios) minus the average return on the two growth portfolios (the low BE/ME ratios), (5) the momentum, and (6) the risk-free interest rate (one-month treasury bill rate). 

\subsection{Fund Performance Measurement}\label{sec.sub:alpha}
\citet{Carhart97} four-factor model has been widely used for measuring a mutual fund's performance. Consider the four-factor model
\begin{equation*}
r_{i,t}= \alpha_i+b_ir_{m,t}+s_ir_{smb,t}+h_ir_{hml,t}+m_ir_{mom,t}+\epsilon_{i,t} \,,
\end{equation*}
where $r_{i,t}$ is the month $t$ excess return of fund $i$ over the risk free rate (proxied by the monthly 30-day T-bill beginning-of-month yield), $r_{m,t}$ is the month $t$ excess return on the CRSP NYSE/NASDAQ value-weighted market portfolio, $r_{smb,t}$, $r_{hml,t}$ and $r_{mom,t}$ are the month $t$ returns on zero-investment factor mimicking portfolios for size, book-to-market, and momentum, and $\epsilon_{i,t}$ is the unobserved random error. Adjusting for the market factors, the intercept $\alpha$ reflects a fund manager's true capability to cover the trading cost and expenses. By convention, the positive $\alpha$ indicates skilled performance.  Skilled fund managers are considered to have stock-picking ability to make extra profits after fees and expenses, while unskilled fund managers' ability is insufficient to cover the trading costs and expenses.

We should point out that there are other asset pricing models in finance. \citet{Jensen68} first introduced Jensen's alpha to measure fund performance, which is the intercept of the CAPM model. The CAPM model directly regresses the excess monthly return on the market factor $r_{m,t}$, and explains fund returns on the overall market.  \cite{FF93} three-factor model further considered $r_{m,t}$, $r_{smb,t}$ and $r_{hml,t}$ as the market factors and has been demonstrated prominent performance in asset pricing. It is worth pointing out that our method does not rely on the assumption that the residuals in the four-factor model should be independent. Thus, misspecification in the model due to a lack of additional common factors does not invalidate our method. Nevertheless, the Carhart four-factor model is sufficient to demonstrate our method for analyzing mutual fund data. Our method can also be applied to the other pricing models according to the readers' preferences. 

The pursuit of positive alpha is based on the rationale that fund managers' skills in investment persist through a certain period of years. Such rationale has been documented in the literature, e.g., \citet{BG95}. Therefore, for a fund manager with a positive value of $\alpha$ in the prior several years, we expect that she would maintain her good quality of investment in the future, at least for a few years. In practice, true $\alpha$'s are unobservable. Instead, we can calculate the ordinary least squares (OLS) estimate of $\alpha$'s for each 10 year period. During a certain period, $t = 1, \cdots, T$, let $\br_i=(r_{i,1},\cdots,r_{i,T})^\top$, $\br_m=(r_{m,1},\cdots,r_{m,T})^\top$, $\br_{smb}=(r_{smb,1},\cdots,r_{smb,T})^\top$, $\br_{hml}=(r_{hml,1},\cdots,r_{hml,T})^\top$, $\br_{mom}=(r_{mom,1},\cdots,r_{mom,T})^\top$, $\bR=(\br_m,\br_{smb},\br_{hml},\br_{mom})$, $\bone=(1,\cdots,1)^\top$. By standard linear model theory, the cross-sectional OLS estimator for $\alpha_i$ can be written as $\widehat{\alpha}_i= \bh^\top\br_i$ where $\bh^\top =  [(T -\bone^\top\bR(\bR^\top\bR)^{-1}\bR^\top\bone)^{-1}\bone^\top-\frac{1}{T}\bone^\top\bR(\bR^\top\bR-\bR^\top\bone\frac{1}{T}\bone^\top\bR)^{-1}\bR^\top]$, for $i = 1, \cdots, p$. The detailed derivation will be given in the Supplementary Materials~A. It is clear that $\cov(\widehat{\alpha}_i,\widehat{\alpha}_j)=\bh^\top \cov(\br_i,\br_j)\bh$, which suggests that there is dependence among $\{\widehat{\alpha}_i\}_{i=1}^p$. 

Based on \citet{Carhart97} four-factor model, with a larger sample size $T$, the OLS estimate of the true $\alpha$ will be more accurate, which encourages us to choose a wider window for the analysis. However, a wider window will possibly raise the issue of survivor bias, that is, funds surviving for a longer time in the past also indicate the success of the funds. Considering such a trade-off, we focus on a 10-year window for applying the \citet{Carhart97} four-factor model, instead of a 5-year window (reduce the estimation accuracy) or a 15-year window (raising survivor bias). 


\subsection{Challenges: Dependence and Nonnormality}\label{sec.sub:dep.nonnormal}
\begin{figure}[!ht]
\centering
\scalebox{0.40}{\includegraphics{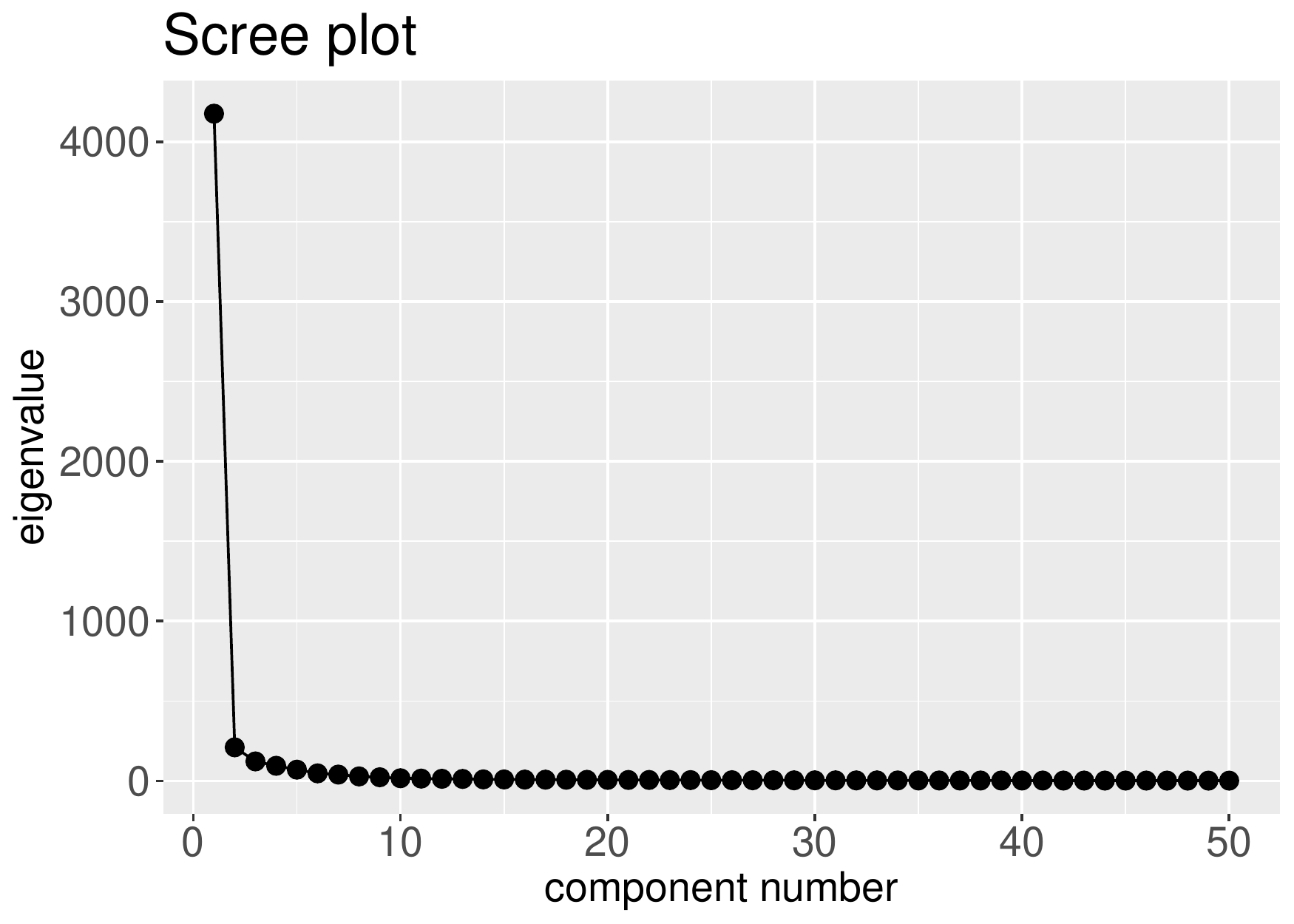}}
\vspace{-0.3cm}
\caption{\footnotesize Top 50 eigenvalues of $\bSigma$ for the mutual fund data between 2009 and 2018.}\label{fig:eig_top_50}
\end{figure}

Based on our derivation in Section~\ref{sec.sub:alpha}, the OLS estimates $\{\widehat{\alpha}_i\}_{i=1}^p$ possess covariance dependence. Let $\bSigma^{\star}$ denote the covariance matrix where the $(i,j)$th element of $\bSigma^{\star}$ is $\bh^\top \cov(\br_i,\br_j)\bh$ and $\cov(\br_i,\br_j)$ is calculated as the sample covariance between the observations $\br_i$ and $\br_j$. We further let $\bSigma$ denote the correlation matrix of $\bSigma^{\star}$. To illustrate this dependence issue, we consider the equity fund data for a 10-year window from 2009 to 2018 as an example. There were $5{,}123$ equity funds that survived during this period. We plot the eigenvalues of $\bSigma$ in Figure~\ref{fig:eig_top_50}. The largest eigenvalue is $4{,}176$, which shows that there is strong dependence. Meanwhile, the largest eigenvalue is $2{,}087$ for $2{,}722$ funds during 2000-2009. The largest eigenvalue increases when the number of funds increases, suggesting that the strong dependence will not be weakened by involving more funds.

\begin{figure}[!ht]
 \centering
\begin{subfigure}[b]{0.38\linewidth}
\centering
 \includegraphics[width=\textwidth]{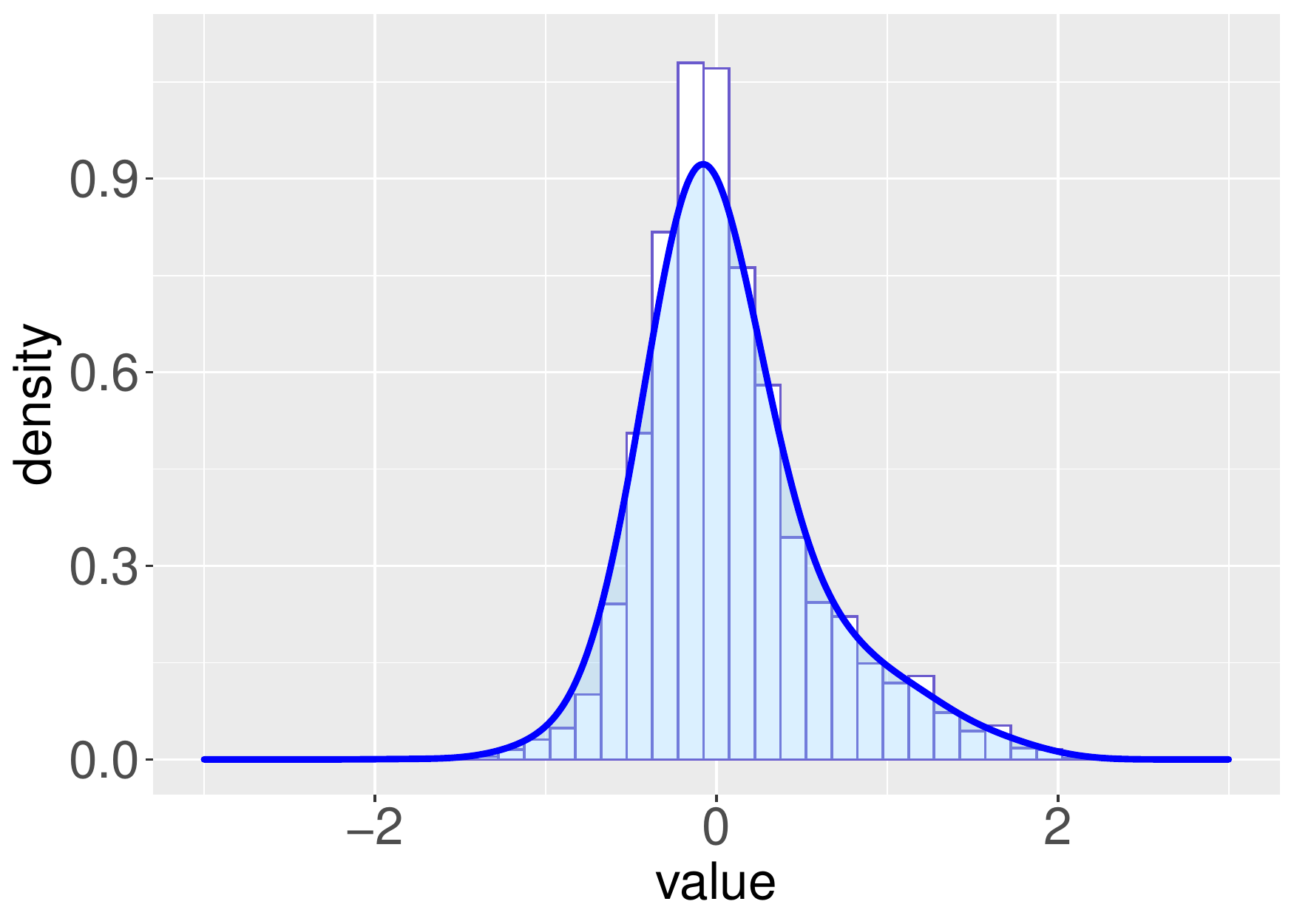}
\vspace{-0.4cm}
\caption{\footnotesize 2002-2011}\label{fig:z_1}
\end{subfigure}
\hspace{1cm}
\begin{subfigure}[b]{0.38\linewidth}
\centering
 \includegraphics[width=\textwidth]{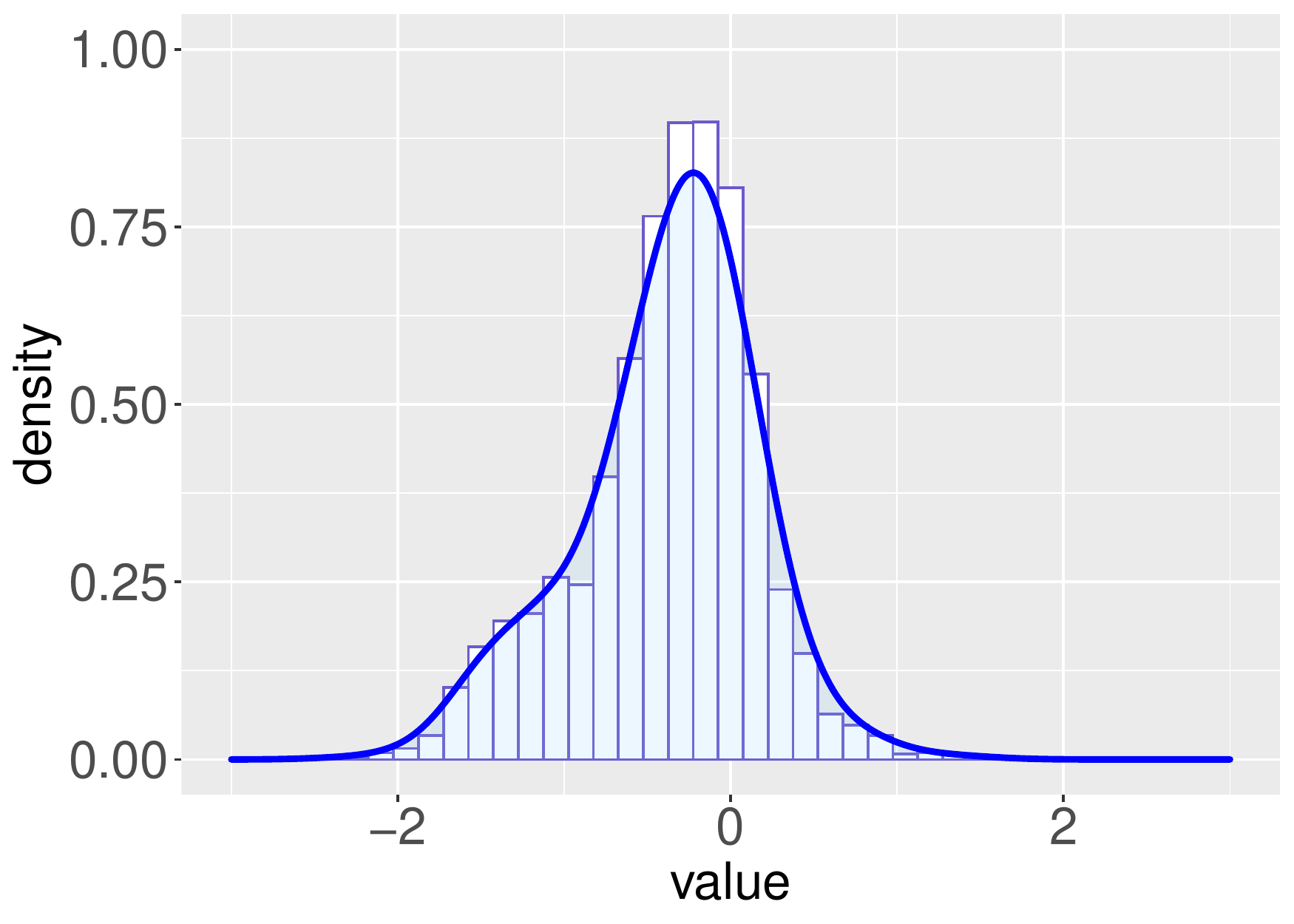}
\vspace{-0.4cm}
\caption{\footnotesize 2009-2018}\label{fig:z_2}
\end{subfigure}
\vspace{-0.2cm}
\caption{\footnotesize Histograms and density plots of standardized OLS estimates for fund $\alpha$'s during (a) 2002-2011 and (b) 2009-2018.}\label{fig:z}
\end{figure}

Various reasons are contributing to this strong dependence. One possible issue is the herding among the mutual funds, which is a common phenomenon documented by the literature \citep{Wermers99}. Herding could arise when mutual funds are applying similar investment strategies. For example, funds often purchase the past year's winners and sell the losers. As a result, these funds herd into (or out of) the same stocks at the same time. 
Another phenomenon we observed is that the empirical distributions of standardized $\{\widehat{\alpha}_i\}_{i=1}^p$ are nonnormal.
\citet{KTWW06} explain the nonnormality phenomenon from two aspects: individual mutual fund $\alpha$'s nonnormality and different levels of risk-taking among funds. The nonnormality arising from distinct levels of risk could be alleviated by standardization, but the normalized statistics are still affected by fund-level nonnormality, caused by many reasons such as nonnormal benchmark returns. 
Figure~\ref{fig:z} plots two distributions of standardized OLS estimates of $\alpha$'s as an example. Figure~\ref{fig:z_1} has a bell shape roughly centered at 0, but with a long tail on the right. In Figure~\ref{fig:z_2}, there is a mode on the left, not completely separated from the centered bell. 

\section{Degree of Non-Skillness (d-value)}\label{sec:conditional}

\subsection{Fund Selection by Multiple Testing Strategy}
Recall that selection of the skilled mutual funds can be formulated as simultaneously testing
\begin{equation}\label{eq.skilled.test}
H_{i0}: \alpha_i\leq0 \quad\quad\text{vs.}\quad\quad H_{ia}: \alpha_i>0, \quad i=1,\cdots,p \,,
\end{equation}
where $p$ is the number of mutual funds during a certain period. Then, the rejection of the $i$th null hypothesis ($\alpha_i \leq 0$) represents the selection of the $i$th fund. Let $\bSigma^{\star}$ be the covariance matrix as calculated in Section \ref{sec.sub:dep.nonnormal}. Let the diagonal elements of $\bSigma^{\star}$ be $\{\sigma_i^2\}_{i=1}^p$. Standardize $\bSigma^{\star}$ to a correlation matrix $\bSigma$. For notational convenience, let $\bZ=(\widehat{\alpha}_1/\sigma_1,\cdots,\widehat{\alpha}_p/\sigma_p)^\top$, where $\bZ$ is the vector of the standardized OLS estimators to avoid heteroscedasticity. 

As shown by Theorem \ref{thm:opt.decision}, the optimal multiple testing procedure can be constructed based on $P(\alpha_i\leq 0|\bZ)$ for $i=1, \cdots, p$, since the result in Theorem \ref{thm:opt.decision} is not restricted to the mean vector. We consider such conditional probabilities, that is, conditional on the data we have, how likely each fund is not skilled. Intuitively, the smaller the value of $P(\alpha_i\leq 0|\bZ)$, the more likely the corresponding $\alpha_i$ is positive. Since we condition on the test statistics from all the mutual funds in our study, we do not lose the dependence information and the nonnormality structure. The construction of this conditional probability reveals our belief that even if we only look at a particular mutual fund, the information of all other funds is important for our understanding of this mutual fund's performance. 

Let $\bZ_{(i)}$ be the vector of elements in $\bZ$ excluding $Z_i$, then $P(\alpha_i\leq0|\bZ)$ can be connected to the local FDR defined by \citet{Efron07}:
\begin{equation}\label{eq:connect_local}
P(\alpha_i\leq0|\bZ)=\frac{f(\bZ_{(i)}|Z_i,\alpha_i\leq0)}{f(\bZ_{(i)}|Z_i)}\times P(\alpha_i\leq 0|Z_i)\,,
\end{equation}
where $f(\bZ_{(i)}|Z_i,\alpha_i\leq0)$ and $f(\bZ_{(i)}|Z_i)$ are the conditional density functions. The term $P(\alpha_i\leq 0|Z_i)$ is the local FDR, where the conditional information is only $Z_i$. When evaluating the performance of the $i$th mutual fund, local FDR only focuses on the data of this fund and ignore the information from other funds. However, the first part  of~\eqref{eq:connect_local} contains the dependence information that makes $P(\alpha_i\leq0|\bZ)$ different from the local FDR. 

The conditional probability $P(\alpha_i\leq 0|\bZ)$ is the posterior probability of the null hypothesis given the test statistics, so it can be viewed as a generalized version of local FDR. In the statistics literature, such posterior probability has been considered as a significance measure where the test statistics possess some special dependence structures, e.g., \cite{SZG08} consider equal correlation, and \cite{TZ07} consider a time series dependence. The references are not exhaustive here. In our mutual fund selection problem, the test statistics possess a more general dependence that has not been well studied in the past. For ease of the presentation, we call $P(\alpha_i\leq0|\bZ)$ the degree of non-skillness (d-value) for the $i$th mutual fund. The calculation of $P(\alpha_i\leq0|\bZ)$ requires delicate analysis, but such efforts will be paid off as shown by our analysis for the mutual fund data in Section~\ref{sec:real}. Further discussions on the d-values will be given in Section~\ref{sec:conditional}. 

With d-values, we can perform the FDR control method (Algorithm~\ref{alg:general}) to select skilled funds. Regarding Corollary \ref{prop:monoton}, the monotonicity property of this procedure has an immediate advantage in the mutual fund selection problem. If we consider two $\theta$ levels: $\theta_1$ and $\theta_2$ with $\theta_1\leq\theta_2$, then the selection set for $\theta_1$ belongs to the selection set for $\theta_2$ due to the monotonicity. Suppose an investor selected skilled funds based on our procedure. She was cautious in the beginning and set $\theta=0.1$. Then, she did further investigations and analysis for the selected funds. Later, she realized that her earlier selection was too conservative, and she would like to update the threshold $\theta$ to 0.15. Fortunately, she does not need to redo the analysis for the funds based on her new selection, because her new selection is only an expansion of her earlier choice. In contrast, the well-known \citet{BH95} procedure does not possess this property. More specifically, the BH procedure rejects the largest $k$ tests such that $k^{-1}p_{(k)}\leq \theta/p$ where $p_{(k)}$ is the $k$th ordered p-value. A larger $k$ might produce a smaller $k^{-1}p_{(k)}$, which violates the monotonicity. 

The multiple testing in (\ref{eq.skilled.test}) is also related to testing inequality. In such a setting, removing the hypothesis with very negative test statistics may enhance the testing power \citep{RW18}. For example, \citet{GLX21} consider a threshold $-\log(\log T)\sqrt{\log p}$ to screen the test statistics. Since the number of hypotheses has been reduced, the sequence of thresholds in their BH procedure would be adjusted to improve the testing power. For each 10-year period in our data, very few test statistics have values smaller than this threshold, so this testing inequality issue has a limited effect on our mutual fund selection problem. More generally, removing a substantial amount of negative test statistics may affect the posterior probabilities through the dependence among the test statistics, but the impact is not clear, which can be pursued in our future research.

\subsection{Mixture Model for Nonnormal and Dependent Test Statistics}\label{sec.sub:mix.mod}
In this section, we propose an approach to calculate the d-value $P(\alpha_i\leq 0|\bZ)$ for the $i$th fund. Let $\mu_i=\alpha_i/\sigma_i$ for $i=1,\cdots,p$, then the signs of $\alpha_i$ and $\mu_i$ are the same, i.e., $P(\alpha_i\leq0|\bZ)=P(\mu_i\leq0|\bZ)$. As a result, we can calculate $P(\mu_i\leq0|\bZ)$ for d-value. 

If the \citet{Carhart97} model fully explains the mutual fund data, in the ideal setting that the random errors are Gaussian distributed with known variances, conditional on $\bmu$, $\bZ$ should follow a multivariate normal distribution. Based on our observation of the empirical distribution of $\bZ$, the nonnormality raises the challenge in modeling the data. This motivates us to propose an appropriate prior for $\bmu$. Recall that Figure~\ref{fig:z_1} has a heavy tail on the right, and Figure~\ref{fig:z_2}  has a bell shape centered roughly at 0 as well as a left mode. The empirical distributions of $\bZ$ for the other 10-year periods are presented in Supplementary Materials~E, which also show the different shapes of nonnormal distributions. We consider a mixture prior to modeling the data structure as follows: 
\begin{eqnarray}\label{eq25}
 \mathbf{Z}|\boldsymbol{\mu} & \sim& N_p(\boldsymbol{\mu} ,\mathbf{\Sigma} )\nonumber\\
 \mu_i &\sim& \pi_0 \delta_ {\nu_0} + \pi_1  N(\nu_1, \tau_1^2)  + \pi_2  N(\nu_2, \tau_2^2) \,, \, i = 1, \cdots \,, p\,,   
\end{eqnarray}
where $\pi_0 + \pi_1 + \pi_2 = 1$ and $\delta_{\nu_0}$ denotes a point mass at $\nu_0$ with $\nu_0 \leq 0$. Note that $\bSigma$ is a known covariance matrix given the data. Each $\mu_i$ is distributed at a point mass $\nu_0$ with probability $\pi_0$ to capture the centered bell shape, and it can be distributed as a normal distribution $N(\nu_1,\tau_1^2)$ with probability $\pi_1$ and a normal distribution $N(\nu_2,\tau_2^2)$ with probability $\pi_2$ to capture the right and the left modes. The empirical distribution of $\bZ$ shows that the point mass $\nu_0$ is close to 0 but not necessarily exact at 0. There are many well-established methods for mixture models with independent $Z_i|\mu_i$, thus the joint likelihood can be decomposed as the product of the marginal likelihoods \citep{RZ11,HL18}. In our problem, such independence assumption is not satisfied. Therefore, the conventional expectation-maximization (EM) algorithm cannot be applied here for fitting the parameters. To the best of our knowledge, estimation under mixture models with such general dependence is still an open question. To facilitate our analysis, we propose a parametric model here, while a more flexible nonparametric modeling technique can be our future research interest.  

\subsection{Fitting Parameters in Mixture Model}\label{sec.sub:AEB}
Recall in the mixture model (\ref{eq25}), there are eight unknown parameters: the proportions $\pi_0$, $\pi_1$ and $\pi_2$, the point mass $\nu_0$, and the parameters $\nu_1$, $\tau_1^2$, $\nu_2$ and $\tau_2^2$ for the two normal distributions. Among these parameters, there are seven free parameters, since we have the restriction $\pi_0 + \pi_1 + \pi_2 = 1$. The degree of non-skillness $P(\mu_i\leq 0|\bZ)$ and the subsequent multiple testing procedure all rely on these parameters which are unknown in practice. We will propose a method called ``approximate empirical Bayes" to estimate these parameters. Several challenging issues arise due to the dependence and the number of parameters. 

Motivated by \citet{FHG12}, we can express $\bZ$ as an approximate factor model. More specifically, let $\lambda_1,\cdots,\lambda_p$ be the non-increasing eigenvalues of $\bSigma$, and $\bgamma_1,\cdots,\bgamma_p$ be the corresponding eigenvectors. For a positive integer $l$, we can write  
\begin{equation}\label{eq6}
 \mathbf{Z} = \boldsymbol{\mu} + \mathbf{CV} + \mathbf{K}\,,
 \end{equation}
where $\mathbf{C} = (\sqrt{\lambda_1}\bgamma_1,\cdots\,, \sqrt{\lambda_{l}}\bgamma_{l})$, $\mathbf{V} \sim N_{l}(0, \mathbf{I}_{l})$ and $\mathbf{K} \sim N_p(0,\mathbf{A})$ with $\mathbf{A} = \sum_{i = l + 1}^p \lambda_i \bgamma_i \bgamma_i^\top$. When $l$ is appropriately chosen, \citet{FHG12} showed that $\bK$ are weakly dependent. Here, we choose $l = \max\{ i \mid \lambda_i > 1\}$, and the weak dependence holds because $\lambda_1$ is extremely large in our data analysis. Let $\mathbf{C} = (\mathbf{c}_1, \cdots\,, \mathbf{c}_p)^\top$ and the diagonal elements of $\mathbf{A}$ be $\eta^2_1,\cdots\,,\eta^2_p$. Then, $\eta ^2_i = 1 - \left\Vert \mathbf{c}_i \right\Vert^2$. Let $ \mathbf{H} = \mathbf{Z} - \mathbf{CV} - \nu_0 \boldsymbol{1}$, then $\mathbf{H} \sim N_p(\boldsymbol{\mu} - \nu_0 \boldsymbol{1},\mathbf{A})$. By location shift,  we have
$$ \mu_i - \nu_0 \sim \pi_0 \delta_0  + \pi_1  N(\nu_1 - \nu_0, \tau_1^2)  + \pi_2  N(\nu_2 - \nu_0, \tau_2^2) \,, \, i = 1, \cdots \,, p\,. $$
Let $u_1 = \nu_1 - \nu_0$ and $u_2 = \nu_2 - \nu_0$. Then, for $ i = 1, \cdots \,, p$, the first four moments of $H_i$ (the $i$th element of $\bH$) are
\begin{align}\label{eq5}
    EH_i & = \pi_1 u_1 + \pi_2 u_2,\nonumber \\
    EH_i^2 &= \pi_1 u_1^2 + \pi_2 u_2^2 + \eta_i^2 +  \pi_1\tau_1^2 + \pi_2\tau_2^2,\nonumber\\
    EH_i^3 &= \pi_1 u_1^3 + \pi_2 u_2^3 + 3(\tau_1^2 + \eta_i^2)\pi_1 u_1 + 3(\tau_2^2 + \eta_i^2)\pi_2 u_2,\\
    EH_i^4 &= \pi_1 u_1^4 + \pi_2 u_2^4 + 6(\tau_1^2 + \eta_i^2)\pi_1 u_1^2 + 6(\tau_2^2 + \eta_i^2) \pi_2 u_2^2\nonumber  \\
    &+ 3(\tau_1^2 + \eta_i^2)^2\pi_1 + 3(\tau_2^2 + \eta_i^2)^2\pi_2 + 3\eta_i^4\pi_0\,.\nonumber
\end{align}
Normally we need to set up seven equations to solve these seven free parameters. However, more equations will increase the difficulty of solving the equation set and higher moments of $H_i$ will also have more complicated expressions of the parameters. Therefore, we only focus on the first four moments of $H_i$ and propose an approximate empirical Bayes method to estimate these parameters. In Algorithm~\ref{alg:para.est}, we present the method in details. We need to point out that Step~\ref{step:solve}, solving the equation set~\eqref{eq5}, is not a trivial task. For brevity, we relegate the detailed derivation steps to the Supplementary Materials~D.  In Step~\ref{step:m.per.cent} of the algorithm, note that $\nu_0$ is very close to 0; thus, smaller values of $|Z_i|$ tend to have $\mu_i$ close to 0 in expression (\ref{eq6}). In Step~\ref{step:moment}, note that $\{H_i\}_{j=1}^p$ are weakly dependent so that the sample moments are expected to converge to the population moments in probability.

\begin{algorithm}[htb!]
\caption{ Parameter Estimation Algorithm 
} \label{alg:para.est}
\SetKw{KwBy}{by}
\SetKwInOut{Input}{Input}\SetKwInOut{Output}{Output}
\SetAlgoLined

\Input{$\bZ$ is the vector of standardized OLS estimators; p is the length of $\bZ$;
$\bC$ is the matrix $(\sqrt{\lambda_1}\gamma_1,\cdots\,, \sqrt{\lambda_{l}}\gamma_{l})$;
$\bSigma$ is the covariance matrix of $\bZ$;
$D_m$ is a search region for $m$ with a increment 5;
$D_{\nu_0}$ is a search region for $\nu_0$ with a increment 0.1;
$D_{( \tau^2_1,\tau^2_2)}$ is a search region for $(\tau^2_1,\tau^2_2) $ with a increment 0.01 for both.
}

$\bZ^{abs} \leftarrow$  $\textsf{sort}(|\bZ|)$ \tcp*{sort the absolute values  $|Z_i|$ in increasing order} \label{step:m.per.cent}

$TV_{comp} \leftarrow 1$

\For{$m$ in $D_m$}{
$index \leftarrow \{i \mid |Z_i| \leq \bZ^{abs}[\textsf{as.integer}(p \times m\%)]\}$

$\bZ^{part} \leftarrow  \bZ[index]$ \tcp*{select $m\%$ $Z_i$ with the smallest absolute values}

$\bC^{part} \leftarrow \bC[index,]$ \tcp*{select  $\mathbf{c}_i$ with respect to the selected $m\%$ $Z_i$ }

$\widehat{\bV} \leftarrow$ the coefficient vector of $\textsf{L1regression}(\bZ^{part} \sim \bC^{part})$ 

\For{$\nu_0$ in $D_{\nu_0}$}{
$\widehat{\bH} \leftarrow \bZ - \bC \widehat{\bV} - \nu_0 \boldsymbol{1}$  \tcp*{ estimate $\bH = \bZ - \bC \bV - \nu_0 \boldsymbol{1}$}

$(EH_i, EH_i^2 ,EH_i^3,EH_i^4) \leftarrow 
({\sum_{i=1}^p\widehat{H}_i}/{p}, {\sum_{i=1}^p\widehat{H}_i^2}/{p}, {\sum_{i=1}^p\widehat{H}_i^3}/{p} ,{\sum_{i=1}^p\widehat{H}_i^4}/{p})$\label{step:moment}

\For{$(\tau_1^2, \tau_2^2)$ in $D_{( \tau^2_1,\tau^2_2)}$}{

$(\widehat{\pi}_0, \widehat{\pi}_1, \widehat{\pi}_2, \widehat{u}_1, \widehat{u}_2) \leftarrow$ solve  \eqref{eq5} based on $(EH_i, EH_i^2, EH_i^3, EH_i^4, \tau_1^2, \tau_2^2)$ \label{step:solve}

$(\widehat{\nu}_1, \widehat{\nu}_2) \leftarrow (\widehat{u}_1 + \nu_0, \widehat{u}_2 + \nu_0)$

$\Tilde{\bZ} \leftarrow $ simulate data following $N_p(\widehat{\boldsymbol{\mu}} ,\mathbf{\Sigma})$ where $\widehat{\mu}_i \sim \widehat{\pi}_0 \delta_ {\nu_0} + \widehat{\pi}_1  N(\widehat{\nu}_1, \tau_1^2)  + \widehat{\pi}_2  N(\widehat{\nu}_2, \tau_2^2) \,, \, i = 1, \cdots \,, p\,,$ 

$TV \leftarrow$ the total variation between $\bZ$ and $\Tilde{\bZ}$

\If{$TV < TV_{comp}$}{
$TV_{comp} \leftarrow TV$ \tcp*{choose the smallest total variation}

$SET_{best} \leftarrow (m, \tau_1^2, \tau_2^2, \nu_0, \widehat{\pi}_0, \widehat{\pi}_1, \widehat{\pi}_2, \widehat{\nu}_1, \widehat{\nu}_2)$  
}
}
}
}
\Output{$SET_{best}$}
\end{algorithm}


We apply this approximate empirical Bayes method to estimate the parameters in the mixture model  for the empirical distribution of $\bZ$ for every 10-year period from 2000-2009 to 2009-2018. Table~\ref{tab1} shows that the total variations between the empirical distribution of $\bZ$ and the distribution of the simulated data based on the fitted mixture model are all below 0.017, suggesting that our fittings capture the nonnormal distribution well.  For the $\nu_1$ and $\nu_2$, it is worth mentioning that we did not restrict $\nu_1\leq\nu_2$ in the fitting. Thus, $\nu_1$ and $\nu_2$ could be the right mode or the left mode. It takes about 3 minutes to fit the parameters for a 10-year period on a Macbook. Time was computed by \texttt{proc.time()} function in R. While our method is more computationally intensive than some testing approaches \citep{BH95, Storey02}, our efforts in such complicated modeling have much better performance, which will be demonstrated in Section~\ref{sec:sim} and ~\ref{sec.sub:return}. 

\begin{table}[!ht]
\caption{\footnotesize Fitted parameters, the proportion of data for $L_1$ regression and the total variation (TV) for 10-year periods.}\label{tab1}
\vspace{-0.3cm}
\begin{center}
\footnotesize 
\begin{tabular}{c|ccccccccccc}
 \hline\hline
         & $m\%$ & $\tau_1^2$ & $\tau_2^2$ & $\nu_0$ & $\pi_0$ & $\pi_1$ & $\pi_2$ & $\nu_1$ & $\nu_2$  & TV  \\
\hline      
2000-2009 &15& .11& .16& .00& .1327& 7.55E-02 & .7918& 1.5412&.2625& .0082\\
2001-2010 & 30&.10&.16& -.20& .1612&  7.00E-02&.7687&1.3997&.0087&.0151\\
2002-2011 &20&.15&.15&-.10&.1552&8.53E-02&.7595&1.1814&.0402&.0067\\
2003-2012 &25&.14&.14&-.20&.2173&6.87E-02& .7139&.7895&-.0935&.0050\\
2004-2013 &25&.15&.15& -.20& .2384& 5.73E-03&.7559&1.2090&-.1341&.0077 \\
2005-2014 &30&.14&.14& -.10&.2605&7.96E-01&.0182&-.2084&.8643&.0092\\
2006-2015 &35&.12&.17&.00& .1471& 2.02E-05& .8529& 1.5978& -.1206&.0054\\
2007-2016  &30& .11&.21&.00&.2381&5.50E-05&.7619&-1.126&-.2885&.0106\\
2008-2017  &40 &.12 &.26 & -.10 &  .2691 &  2.29E-04 & .7307 & -1.0716 & -.4247 & .0115\\
2009-2018  &30& .12& .15& -.10& .0759&1.21E-01&.8035&-1.2035& -.2418 &.0169\\
\hline\hline
\end{tabular}
\end{center}
\end{table}

\begin{figure}[!ht]
\caption{\footnotesize Histograms and density plots of $\bZ $ and the simulated data based on the parameter-fitting results from our approximate empirical Bayes method (AEB) during (a) 2002-2011 and (b) 2009-2018.}\label{fig:hist_09_18}
     \centering
     \begin{subfigure}[b]{0.48\textwidth}
    \centering
    \includegraphics[width=\textwidth]{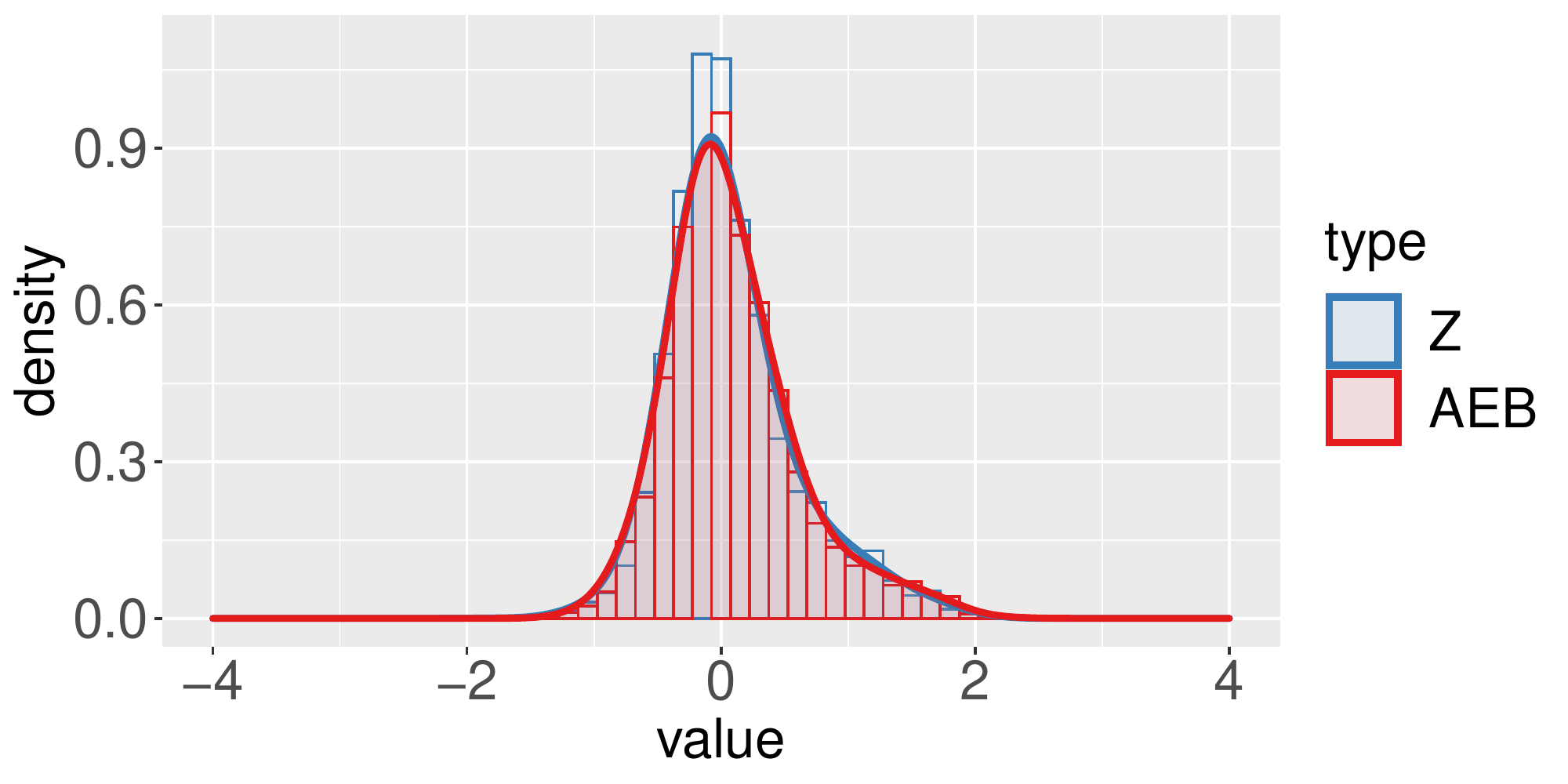}
    \vspace{-0.5cm}
    \caption{\footnotesize 2002-2011}
     \end{subfigure}
     \hfill
     \begin{subfigure}[b]{0.48\textwidth}
         \centering
    \includegraphics[width=\textwidth]{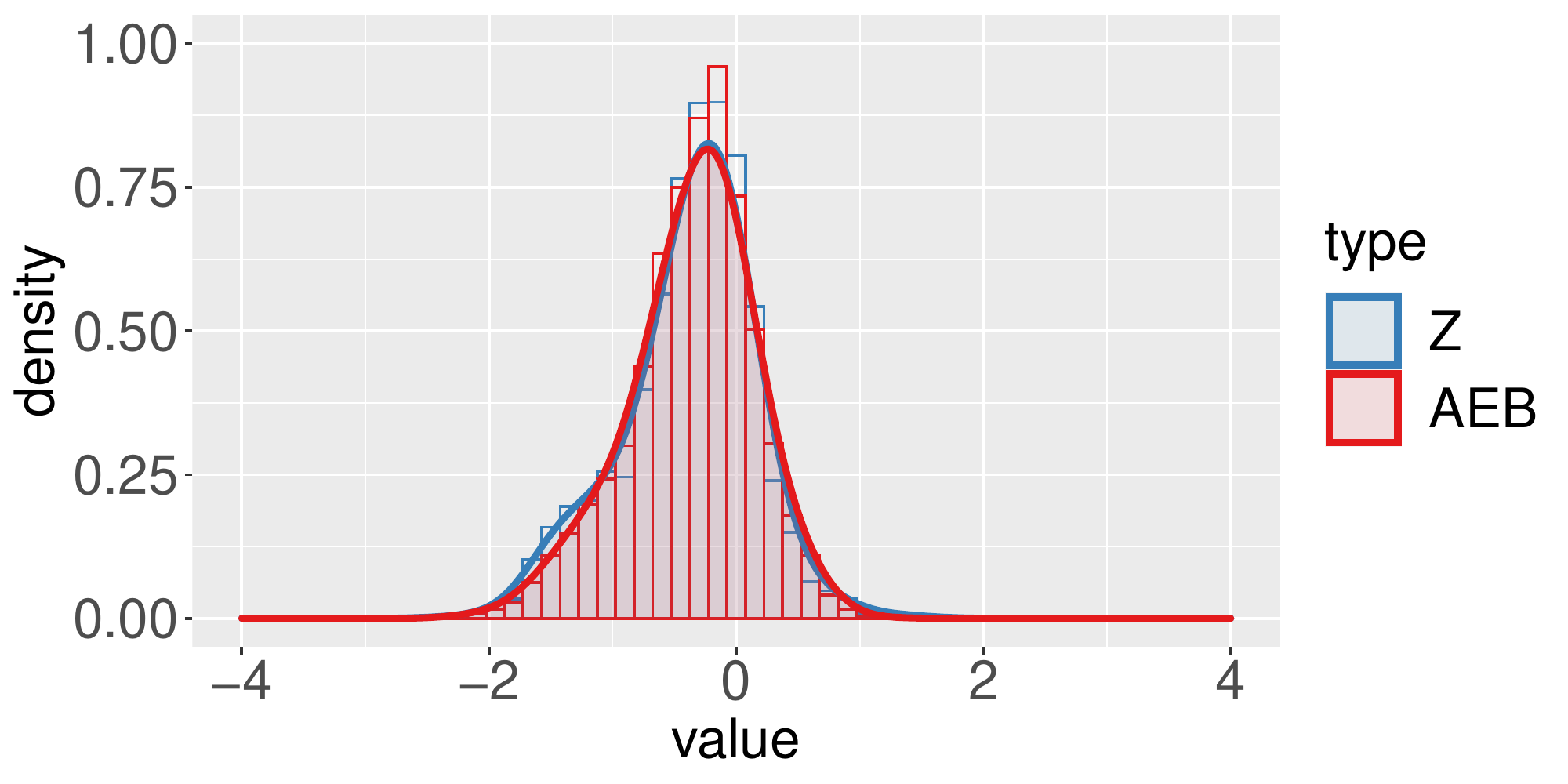}
    \vspace{-0.5cm}
     \caption{\footnotesize 2009-2018}
     \end{subfigure}
     \hfill
\end{figure}

Figure~\ref{fig:hist_09_18} compares the shape of the empirical distribution of $\bZ$ and the simulated data based on the fitted parameters during 2002-2011 and 2009-2018. It is clear that the nonnormality phenomenon exists, so the p-value calculation based on the normality assumption is not sufficient  here. Our proposed mixture model, on the other hand, can capture this structure. Figures~S1 and~S2 (Supplementary Materials~E) demonstrate that the mixture model also well captures the various distributions for different 10-year periods. In Supplementary Materials~E, we also compare our performance with another parameter estimation method developed in \citet{HL18}.

By looking at the empirical distributions, a natural question is whether we should select the funds localized on the right tail part. From a p-value perspective, mutual funds on the right tail are more extreme observations and should be selected as significant. However, d-values for our testing procedure are quite different. The funds on the right tail are not necessarily selected by our procedure, and the funds on the right tail do not necessarily have a persistent performance for the subsequent years. We will further demonstrate this issue in Section~\ref{sec.sub:comp.p.p} with more details. 

\subsection{Calculating Degree of Non-Skillness}\label{sec.sub:cond.prob}
Following the notation in Section~\ref{sec.sub:AEB}, let  $\lambda_1,\cdots,\lambda_p$ be the non-increasing eigenvalues for the covariance  matrix $\bSigma$, and $\bgamma_1,\cdots,\bgamma_p$ be the corresponding eigenvectors. For a positive definite matrix $\bSigma$, the smallest eigenvalue $\lambda_p$ is positive. This helps us construct a strict factor model expression for the test statistics. More specifically, $\bSigma=\sum_{i=1}^p\lambda_i\bgamma_i\bgamma_i^\top. $ We re-arrange this summation as 
\begin{equation*}
\bSigma=\sum_{i=1}^p(\lambda_i-\lambda_p)\bgamma_i\bgamma_i^\top+\lambda_p\sum_{i=1}^p\bgamma_i\bgamma_i^\top=\sum_{i=1}^p(\lambda_i-\lambda_p)\bgamma_i\bgamma_i^\top+\lambda_p\bI. 
\end{equation*}
Therefore, the original data $\bZ$ can be stochastically expressed as follows:
\begin{equation}\label{eq10}
\bZ=\bmu+\bB\bW+\bxi,
\end{equation}
where $\bB=(\sqrt{\lambda_1-\lambda_p}\bgamma_1,\cdots,\sqrt{\lambda_{p-1}-\lambda_p}\bgamma_{p-1})$, $\bW\sim N_{p-1}(0,\bI_{p-1})$ and $\bxi\sim N_p(0,\lambda_p\bI_p)$. Note that if $\lambda_{j}=\lambda_p$, which corresponds to the spiked covariance model, the dimension of $\bW$ will be $(j-1)$, and the subsequent calculation will be greatly simplified. The expression in~\eqref{eq10} is different from expression~\eqref{eq6} in Section~\ref{sec.sub:mix.mod}. Here the elements in $\bxi$ are independent, while the random errors in~\eqref{eq6} are weakly dependent. Consider the row vector of $\bB$ as $\bb_i$ for $i=1,\cdots,p$. For notational convenience, we use $dN(\mu,\sigma^2)$ to denote the probability density function of $N(\mu,\sigma^2)$. The density of $\bZ$ can be expressed as
\begin{multline}\label{eq21}
f(\bZ)=
E_W\{\prod^p_{i = 1}[\pi_0 dN(\nu_0 + \mathbf{b}_i\mathbf{W},\lambda_p) +  \pi_1 dN(\nu_1 + \mathbf{b}_i\mathbf{W}, \tau_1^2 + \lambda_p)  \\
 + \pi_2 dN(\nu_2 + \mathbf{b}_i\mathbf{W}, \tau_2^2 + \lambda_p)] \} \,.
\end{multline}
The last expectation is with respect to $\bW\sim N_{p-1}(0,\bI_{p-1})$. The detailed derivation is in the Supplementary Materials~C. In the expression for $f(\bZ)$, $dN(\nu_0+\bb_i\bW,\lambda_p)$ denotes the density of $N(\nu_0+\bb_i\bW,\lambda_p)$ at point $Z_i$, and the other notations have similar meanings. 

Similarly, $\int_{\mu_i\leq0}f(\textbf{Z}, \mu_i)d\mu_i$ can be expressed as
\begin{multline} \label{eq11}
    \int_{\mu_i\leq0}f(\textbf{Z},\mu_i)d\mu_i =  \, E_W \left\{\prod^p_{j = 1}[\pi_0 dN(\nu_0 + \textbf{b}_j \textbf{W},\lambda_p) \right.\\
    + \pi_1 dN(\nu_1 + \mathbf{b}_j\mathbf{W}, \tau_1^2 + \lambda_p) + \pi_2 dN(\nu_2 + \mathbf{b}_j\mathbf{W}, \tau_2^2 + \lambda_p)] \\ 
     \left. \times \frac{\pi_0 dN(\nu_0 + \textbf{b}_i \textbf{W},\lambda_p) + \pi_1 G_{\tau_1^2, \lambda_p}(\nu_1, \textbf{b}_i \textbf{W}, Z_i) + \pi_2 G_{\tau_2^2, \lambda_p}(\nu_2,\textbf{b}_i \textbf{W}, Z_i)}  {\pi_0 dN(\nu_0 + \textbf{b}_i \textbf{W},\lambda_p) + \pi_1 dN(\nu_1 + \mathbf{b}_i\mathbf{W}, \tau_1^2 + \lambda_p) + \pi_2 dN(\nu_2 + \mathbf{b}_i\mathbf{W}, \tau_2^2 + \lambda_p)}  \right\}\,,
\end{multline}
where
\begin{equation}\label{eq4}
G_{\tau^2, \lambda_p}(\mu_0, \textbf{b}_i \textbf{W}, Z_i)=\frac{1}{\sqrt{2\pi(\lambda_p +\tau^2)}}\exp\left\{\frac{\beta_i^2}{2\sigma^2}-\frac{(Z_i -\textbf{b}_i \textbf{W})^2}{2\lambda_p} - \frac{\mu_0^2}{2\tau^2}\right\}\Phi\left(\frac{- \beta_i}{\sigma}\right)\,.
\end{equation}
The detailed derivation of expression~\eqref{eq4} will be given in the Supplementary Materials~C. Correspondingly, the degree of non-skillness $P(\mu_i  \leq 0 \mid \mathbf{Z}) =\int_{\mu_i\leq0}f(\textbf{Z}, \mu_i)d\mu_i/f(\textbf{Z})$.

So far, we have focused on selecting the skilled funds. Identifying unskilled funds is also important in practice, because investors can possibly avoid substantial loss by such selection. Similar to the multiple testing \eqref{eq.skilled.test}, we can also test $H_{0i}: \,\alpha_i \geq 0$ vs. $H_{ai}:\, \alpha_i < 0\,,$ for $i = 1\,,\,\cdots \,,\,p$. We compute $P(\alpha_i\geq0|\bZ)=P(\mu_i  \geq 0 \mid \mathbf{Z}) =\int_{\mu_i\geq0} f(\textbf{Z}, \mu_i)d\mu_i/f(\textbf{Z})$. Smaller value of $P(\alpha_i\geq0|\bZ)$ indicates that $\alpha_i$ is more likely to be negative. We sort $P(\alpha_i\geq0|\bZ)$ from the smallest to the largest, and denote $\mbox{LOS}_{(1)},\cdots,\mbox{LOS}_{(p)}$ as the non-decreasing order of $P(\alpha_i\geq0|\bZ)$. Given a $\theta$ level, we choose the largest $l$ such that 
\begin{equation}\label{eq:uskilled.test}
l^{-1}\sum_{i=1}^l\mbox{LOS}_{(i)}\leq\theta. 
\end{equation}
For the expression of $\int_{\mu_i\geq0}f(\bZ, \mu_i)d\mu_i$, we have
\begin{align*}
     &\int_{\mu_i\geq0}f(\bZ, \mu_i)d\mu_i  \\
      = & \, E_W \left\{\prod^p_{j = 1}\left[\pi_0 dN(\nu_0 + \textbf{b}_j \textbf{W},\lambda_p) + \pi_1 dN(\nu_1 + \mathbf{b}_i\mathbf{W}, \tau_1^2 + \lambda_p) + \pi_2 dN(\nu_2 + \mathbf{b}_i\mathbf{W}, \tau_2^2 + \lambda_p)\right]\right.\\
     & \left. \times \frac{\boldsymbol{1}\{\nu_0 = 0\}\pi_0 dN(\nu_0 + \textbf{b}_i \textbf{W},\lambda_p) + \pi_1 Q_{\tau_1^2, \lambda_p}(\nu_1, \textbf{b}_i \textbf{W}, Z_i) + \pi_2 Q_{\tau_2^2, \lambda_p}(\nu_2,\textbf{b}_i \textbf{W}, Z_i)}  {\pi_0 dN(\nu_0 + \textbf{b}_i \textbf{W},\lambda_p) + \pi_1 dN(\nu_1 + \mathbf{b}_i\mathbf{W}, \tau_1^2 + \lambda_p) + \pi_2 dN(\nu_2 + \mathbf{b}_i\mathbf{W}, \tau_2^2 + \lambda_p)}  \right\}\,,
\end{align*}
in which
\begin{equation*}
Q_{\tau^2, \lambda_p}(\mu_0, \textbf{b}_i \textbf{W}, Z_i)  = \frac{1}{\sqrt{2\pi(\lambda_p +\tau^2)}}\exp\left\{\frac{\beta_i^2}{2\sigma^2}-\frac{(Z_i -\textbf{b}_i \textbf{W})^2}{2\lambda_p} - \frac{\mu_0^2}{2\tau^2}\right\}\left(1 - \Phi\left(\frac{- \beta_i}{\sigma}\right)\right)\,.
\end{equation*}
The detailed derivation will be given in the Supplementary Materials~C.

\subsection{Comparison with Other Existing Methods}
Our approach is different from the existing methods in the mutual fund literature. We will compare with the following methods concretely: \citet{FC21}: FC; \citet{HL20}: HL20; \citet{FHG12}: FHG; \citet{LD19}: FAT; \citet{GLX21}: GLX; \citet{HL18}: HL18. 

Both FC and HL20 consider type II error in distinct ways from ours. FC improves estimates for the proportions of unskilled, zero-alpha, and skilled funds developed in \cite{BSW10} by considering the power of the test and the confusion parameter, which increases the power of the test. It is a p-value based approach and ignores the dependence. HL20 uses a double-bootstrapped procedure to estimate the FDR and FNR. This approach can be used to compare FNR's of different testing procedures if these testing procedures are provided, but it does not offer an optimal testing procedure. 

FAT and GLX both assume an approximate factor model, in which conditional on the common factors, the random errors have a weakly dependent correlation matrix. By adjusting the common factors and standardization with the standard deviation of the random errors, the signal-to-noise ratio can be increased, thus enhancing the testing power. Since the resulting test statistics are nearly independent, FAT further considers Storey procedure \citep{Storey02}, while GLX considers BH procedure \citep{BH95}. In practice, the fund returns may not satisfy such an approximate factor model, and the random errors may not have a weakly dependent correlation matrix, i.e., by adjusting the common factors and standardization, the resulting test statistics can still have general and strong dependence. BH and Storey procedure will be conservative for such settings. As a related but different approach, FHG directly estimates the FDP by incorporating the strong dependence. It does not rely on the assumption of the approximate factor model. The drawback is that it was designed for the setting with very sparse signals. When there is a substantial proportion of signals, FHG will be conservative. All three methods are p-value based approaches, which are not optimal. Our optimal procedure does not rely on the approximate factor model assumption, and it also adapts to different sparsity of signals. 

HL18 is an estimation approach different from the multiple testing framework. It utilizes a two-part mixture model for the $\alpha$'s, and develops an EM algorithm to estimate $\alpha$'s. The EM algorithm is implemented under the assumption that conditional on the common factors, the random errors are independent. As an intermediate step in our approach, we considered a three-part mixture model for the test statistics, in which the dependence is general. Since it violates the independence assumption in HL18, we develop the approximate empirical Bayes method for fitting the parameters. We will demonstrate through numerical studies that our approach outperforms HL18 in terms of density fitting.

\section{Simulation Studies}\label{sec:sim}

In the simulation studies, we consider $p = 1000$. For each simulation round, we first sample 1000 funds without replacement from the 2009-2018 dataset. The $i$th item in the sample is fund $s(i)$ from the original data. Then, we construct the new returns $\Tilde{r}_{it}$ as
\begin{equation*}
\Tilde{r}_{it} =  \mu_i \sigma_{s(i)} + R_t  \widehat{\beta}_{s(i)} + \title{\epsilon}_{it}\,, \mbox{ for } t = 1, \ldots, T,     
\end{equation*}
where $R_t=(r_{m,t},r_{smb,t},r_{hml,t},r_{mom,t})$, $\widehat{\beta}_{s(i)}$ is the corresponding OLS coefficient vector for fund $s(i)$, and $\sigma^2_{s(i)}$ is the $s(i)$th diagonal element in the correlation matrix for the OLS estimates of $\alpha$, which is discussed in Secion~\ref{sec.sub:dep.nonnormal}. We construct $\mu_i$ according to the mixture model ~\eqref{eq25} with $\pi_0 = 0.1, \nu_0 = 0, \nu_1 = -0.5, \nu_2 = 1.2, \tau_1^2 = 0.1, \tau_2^2 = 0.1$. We consider $\title{\bepsilon}_t  \sim  N_p(0 ,\bSigma_\epsilon)$ where $\title{\bepsilon}_t = (\epsilon_{1,t}, \cdots, \epsilon_{p,t})^\top$. Note that $\pi_1, \pi_2$ and $\bSigma_\epsilon$ will affect the proportion of skilled funds (sparsity) and the dependence structure, respectively. We consider the following settings. 

\noindent \textbf{Sparsity}: The proportion of skilled funds (sparsity) is mainly determined by $\pi_2$, since we set $\nu_2 > 0$. We change the value of $\pi_2$ to achieve different levels of sparsity. Correspondingly, $\pi_1$ is adjusted to ensure that $\pi_0 + \pi_1 + \pi_2 = 1$. Two settings are considered as follows.
\vspace{-0.5cm}
\begin{enumerate}[label= s.\arabic*]
    \item\label{sp.1}  $ \pi_1 = 0.7, \pi_2 = 0.2$.
    \item\label{sp.2}  $\pi_1 = 0.2, \pi_2 = 0.7$.
\end{enumerate}
\vspace{-0.5cm}
\ref{sp.2} produces a denser signal setting than \ref{sp.1}. Some testing procedures, such as FHG, are designed for sparse signals, so their procedures are  more conservative under \ref{sp.2}. 

\noindent \textbf{Dependence:}
Let $\bSigma_\epsilon = \mbox{cor}(\bSigma^\star_\epsilon)$ and $\mbox{diag}(\bSigma^\star_\epsilon) = (\sigma^2_{\bepsilon_{s(1)}}, \ldots, \sigma^2_{\bepsilon_{s(p)}})$, where $\sigma^2_{\bepsilon_{s(i)}} = \sum_t \widehat{\epsilon}^2_{s(i),t} /(T - 1) $ and $\widehat{\epsilon}_{s(i),t} = r_{s(i)} - \widehat{\alpha}_{s(i)} - R_t \widehat{\beta}_{s(i)} $. Three dependence structures are considered as follows.
\vspace{-0.5cm}
\begin{enumerate}[label=d.\arabic*]
    \item\label{d.1}  Strict factor model: $\bA \in \mathbb{R}^{p\times 4}$,  $ \bA_{ij} \sim  N(0,4)$, and $\bSigma_\epsilon =  cor(\bA \bA^\top + \mathbf{I}_p)$.
    \item\label{d.2}  Power-decay model: $\bA \in \mathbb{R}^{p\times 10}$, $ \bA_{ij} \sim  N(0,4)$ and $\bSigma_\epsilon =  cor(\bA \bA^\top + \bM)$ where the $(i,j)$th element in $\bM$ is $M_{ij} = \rho^{|i - j|}$ for $i,j = 1, \cdots, p$ with  $\rho = 0.8$.
    \item\label{d.3}  Long memory autocovariance model: $\bA \in \mathbb{R}^{p\times 10}$, $ \bA_{ij} \sim  \mbox{Uniform}(-1,1)$ and $\bSigma_\epsilon =  cor(\bA \bA^\top + \bM)$ where   $M_{ij} =  0.5||i - j| + 1|^{2H} - 2|i -j|^{2H} + ||i - j| - 1|^{2H}$ for $i,j = 1, \cdots, p$ and $H = 0.9$.
\end{enumerate}
\vspace{-0.5cm}
For the three settings of the dependence, conditional on the four market factors, the random errors possess additional strong dependence. The structure of $\bM$ in \ref{d.3} has also been considered in \citet{BL08}.

\begin{table}[h!!!]
\centering
  \footnotesize
\begin{tabular}{c|c|c|c|c|c|c|c|c|c|c}
\hline\hline
Dependence & Sparsity & Mean & Ours & Storey & BH & FAT & GLX & FC & HL20& FHG \\\hline
    \multirow{6}{*}{\ref{d.1}} & \multirow{3}{*}{\ref{sp.1}} & FDP  & 0.093 & 0    & 0    & 0.162 & 0.043 & 0.061 & 0.020 & 0.068\\
                                        & & FNP  & 0.049 & 0.238   & 0.238 & 0.035 & 0.066 & 0.068 &0.087 & 0.073\\
                             & &  selection & 222   & 0 & 0  & 252 & 195 & 197 & 170 & 192\\\cline{2-11}
                            &\multirow{3}{*}{\ref{sp.2}}   & FDP  & 0.097 & 0.009 & 0  & 0.003 & 0.025 & 0.082 &0.026 & 0.014 \\
                                                & & FNP  & 0.033 & 0.311 & 0.713 & 0.686 & 0.158 & 0.064 & 0.172 & 0.276\\
                                & & selection & 783 & 535 & 0 & 31 & 678 & 760 &674 &603 \\\hline\hline
\multirow{6}{*}{\ref{d.2}} & \multirow{3}{*}{\ref{sp.1}} & FDP  & 0.095 & 0       & 0 & 0 & 0.050 & 0.057 & 0.019 & 0.073  \\
                                         & & FNP  & 0.060 & 0.238   & 0.238 & 0.239 & 0.079 & 0.070 & 0.089 & 0.073 \\
                              & & selection & 212   & 0 & 0  & 0 & 185 & 193 & 168 & 194 \\\cline{2-11}
                            &\multirow{3}{*}{\ref{sp.2}}   & FDP  & 0.101 & 0.006 & 0 & 0 & 0.023 & 0.093 & 0.027 & 0.013\\
                                                & & FNP  & 0.054 & 0.426 & 0.711 & 0.711 & 0.223 & 0.058 & 0.178 & 0.277\\
                                     & &  selection & 778   & 423   & 0  & 0 & 644 & 769 &  669 & 601\\\hline\hline
\multirow{6}{*}{\ref{d.3}} & \multirow{3}{*}{\ref{sp.1}} & FDP  & 0.100  & 0 &  0        &  0.090 & 0.043 &  0.054& 0.016 & 0.055\\
                                            & & FNP  & 0.060  &0.238 & 0.238  &  0.167 & 0.083 & 0.068&0.090 & 0.082\\
                                      & &  selection & 213  & 0&0&  101   &  179& 195 & 167& 181\\\cline{2-11}
                  &\multirow{3}{*}{\ref{sp.2}}   & FDP  & 0.100  & 0.029 & 0 & 0 & 0.020 & 0.064 & 0.026& 0.013\\
                                        & & FNP  & 0.066 & 0.332  & 0.711 & 0.562 & 0.232 &  0.085 &0.173 & 0.303\\
                                   & & selection & 770   & 496    & 0& 291 & 636 & 735 &670 & 566\\\hline
\hline

\end{tabular}  
\caption{\footnotesize Comparison of our approach (Algorithm~\ref{alg:general}) with approximate empirical Bayes method for fitting the parameters and other FDR control methods  for each combination of sparsity (\ref{sp.1}, \ref{sp.2}) and dependence structures (\ref{d.1}, \ref{d.2}, \ref{d.3}) where $\theta = 0.1$. The average false discovery proportion (FDP), false non-discovery proportion (FNP), number of selection are calculated over 100 simulations for each setting. } \label{table:sim_fdr}
\end{table}

We apply our multiple testing procedure (Algorithm~\ref{alg:general}) for each combination of sparsity and dependence with $\theta = 0.1$. As we consider the parameters in the mixture model unknown, our procedure first applies the approximate empirical Bayes method to fit the model before performing the FDR control method. The average FDP, FNP, and number of selection are calculated over 100 simulations in Table~\ref{table:sim_fdr}, and the total variation for the model fitting are presented in Supplementary Materials~F. Furthermore, we compare with other existing FDR control methods \citep{BH95, Storey02, FHG12, FC21, LD19, HL20, GLX21} and model fitting procedure \citep{HL18} in these tables. (Correspondingly, we abbreviate these methods as BH, Storey, FHG, FC, FAT, HL20, GLX and HL18 method.) Table~\ref{table:sim_fdr} illustrates that the average FDP of our approach is closer to 0.1 while the average FNP is smaller than the other methods under various sparsity and dependence structures. This is consistent with our theoretical result that our procedure can minimize FNR while controlling FDR at the desired level.


\section{Skilled Funds Selection with FDR Control}\label{sec:real}
We apply the multiple testing procedure (Algorithm~\ref{alg:general}) utilizing the degrees of non-skillness (d-value) to our downloaded data for equity funds in this section. It shows that our procedure selects funds with persistent performance and produces substantial returns in the subsequent years. The discrepancies in selections compared with the conventional multiple testing procedures mainly result from the differences between p-values and d-values. We compare these two measurements of significance in Section~\ref{sec.sub:comp.p.p} to illustrate the advantage of d-values. 

\subsection{Dynamic Trading Strategy}\label{sec.sub:return}

\begin{figure}[h!!!]
\caption{\footnotesize The values of portfolios selected by our procedure and other FDR control procedures ( BH, Storey, FHG, FC, FAT, HL20 and GLX) from the end of 2009 to the end of 2019. ``Skilled funds" refers to our portfolio. ``S\&P 500" is the value of S\&P 500 index. For ``HL18", we construct the portfolio containing funds with HL18 estimation of alphas greater than its 95\% quantile.}\label{fig:return}
\centering
\scalebox{0.2}{\includegraphics{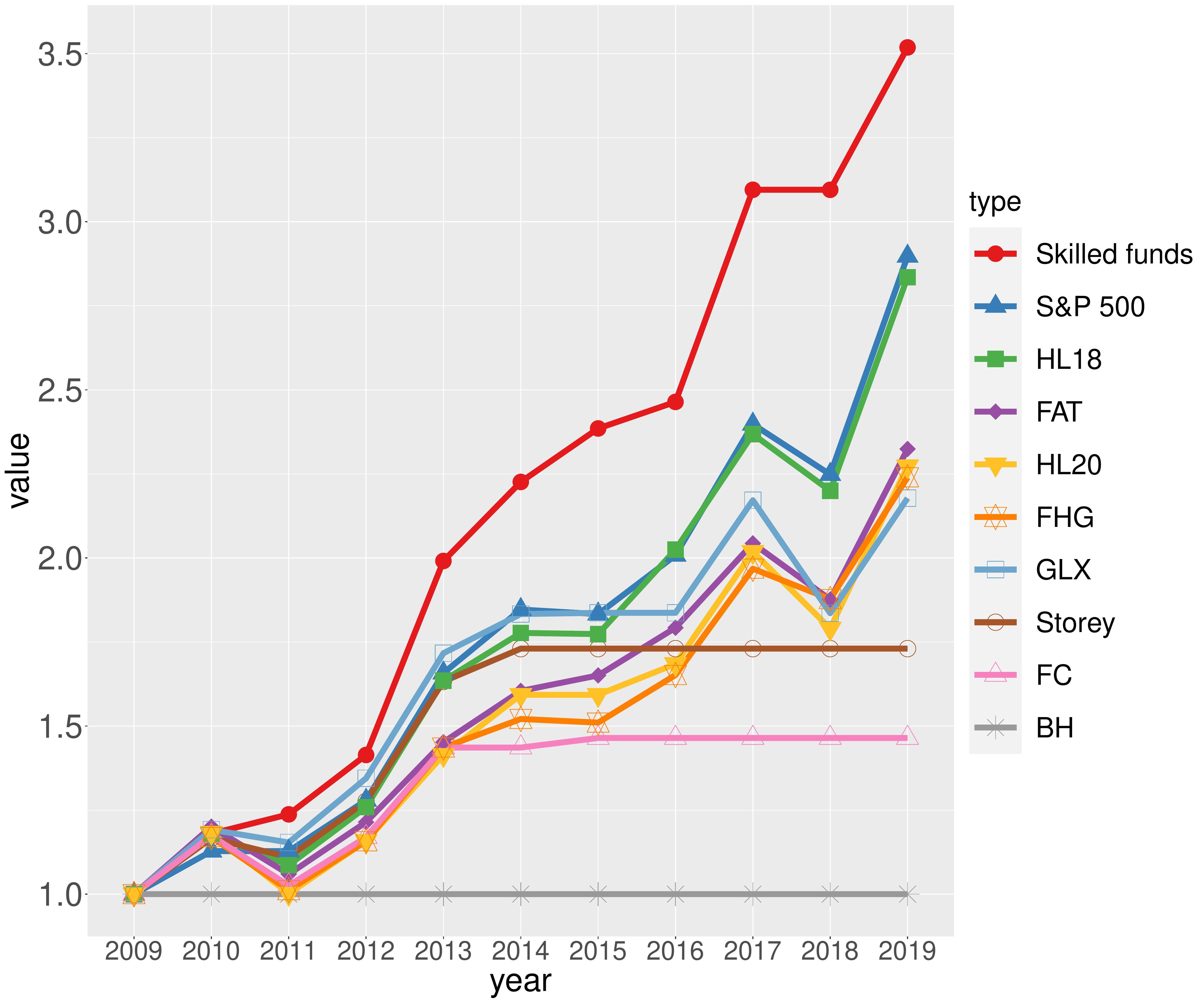}}
\end{figure}

We first demonstrate the short-term out-of-sample performance of our procedure. Hypothetically we had \$1 million at the end of 2009. We selected the skilled funds by Algorithm~\ref{alg:general} with $\theta$ level as 0.15 based on the data from 2000 to 2009. Then we invested in the selected funds with equal weights for the year 2010. At the end of 2010, we re-selected the skilled funds based on the data from 2001 to 2010 and invested in the newly selected funds with equal weights for the year 2011. We continue the above trading strategy until the end of 2019. The value of our portfolio at the end of each year is plotted with a red line in Figure~\ref{fig:return}. For comparison, we first consider two conventional multiple testing procedures for selecting funds: BH procedure and Storey procedure. BH procedure is constructed based on the individual p-values. When there is general dependence among the test statistics, it is usually very conservative. Storey procedure is less conservative than BH procedure, and it can asymptotically control the FDR under the weak dependence of the test statistics. \citet{BSW10} selected skilled funds based on Storey procedure. For a fair comparison, we choose the threshold as 0.15 for BH procedure and Storey procedure. Table~\ref{table:return} shows that after 10 years, our portfolio grows to \$3.52 million (annual return $13.41\%$) while their portfolios are less than \$1.75 million. Due to the conservativeness, both BH procedure and Storey procedure fail to select any funds for many 10-year periods, resulting in the flat lines in Figure~\ref{fig:return}. It is also worth plotting the growth of the portfolio by investing in the S\&P 500 index in Figure~\ref{fig:return}, since S\&P 500 is a common benchmark for financial investment. After 10 years, the S\&P 500 index would grow to \$2.90 million (annual return $11.22\%$), still much inferior to our constructed portfolio. Furthermore, we construct similar equal-weighted portfolios based on the funds selected by the existing methods discussed in Section~\ref{sec:sim}. In particular, we consider FHG, FC, FAT, HL20, GLX, and HL18. Recall that HL18 is not an FDR control method, so funds whose HL18 estimates of alphas greater than 95\% quantile are selected. The rest of the FDR control procedures above are implemented with $\theta=0.15$. The returns of these portfolios are recorded in Figure~\ref{fig:return} and Table~\ref{table:return}, which shows that it is difficult to outperform S\&P 500, while our procedure succeeds.

\begin{table}[h!!!]
\caption{\footnotesize The values of portfolios by the end of 2019 and the annual returns for each approach. }\label{table:return}
\centering
\footnotesize
\begin{tabular}{c|c|c}
\hline\hline
Portfolio & Value (million) & Annual Return (\%) \\ \hline
Our method & 3.52 & 13.41 \\
S\&P 500 & 2.90 & 11.22\\
HL18 & 2.83 & 10.98\\
FAT & 2.32 & 8.80\\
HL20 & 2.27 & 8.56\\
FHG & 2.24 & 8.40 \\
GLX & 2.18 & 8.09\\
Storey & 1.73& 5.64\\
FC & 1.46 & 3.89\\
BH & 1.00 & 0.00\\\hline
\end{tabular}
\end{table}

\subsection{Long-term Performance Measured in Moving 10-year Windows}\label{sec.sub:window}

We further want to evaluate the long-term performance of the skilled funds selected by our procedure. Based on a 10-year window of historical data, we apply our proposed multiple testing procedure to select the skilled funds and the unskilled funds, with the rest considered as the non-selected group. For selecting the unskilled funds by the multiple testing procedure \eqref{eq:uskilled.test}, we set the $\theta$ level as 0.05. The selection of skilled funds is the same as previous section. Remark that in Table \ref{table:exp.turn} none of the funds were selected as skilled funds by our testing procedure for the period 2008-2017, which seems to be a very cautious selection. However, relating to the results in Section~\ref{sec.sub:return} for the dynamic trading strategy, this zero selection leads to zero loss in the year 2018, while S\&P 500 lost 6.24\% in that year. 

\begin{figure}[!ht]
\caption{\footnotesize The medians of estimated annual $\alpha$'s of the three groups selected based on the data from 2003-2012, and the estimated annual $\alpha$'s are calculated for each 10-year window after this period.  }\label{fig:alpha_example}
\vspace{-0.5cm}
\begin{center}
\scalebox{0.2}{\includegraphics{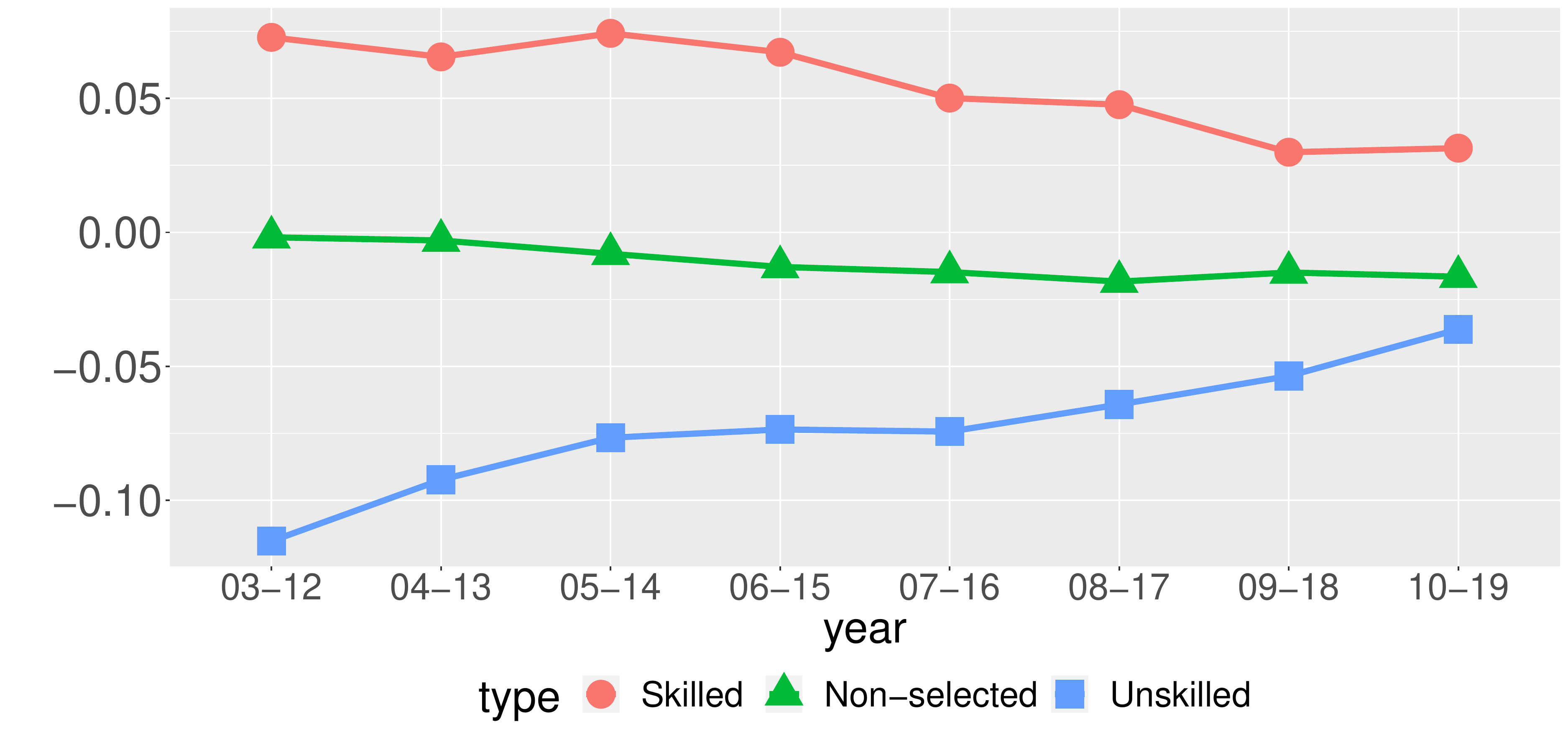}}
\end{center}
\end{figure}

We will consider evaluating the performance of our selected funds by annual $\alpha$ medians. Recall that the $\alpha$ is used to measure the fund manager's capability by adjusting the market factors. To avoid the effect of some outliers, we look at the median of the estimated $\alpha$'s for a selected group. An example is presented in Figure~\ref{fig:alpha_example} where we classify the mutual funds into three groups (skilled, unskilled, non-selected) based on the data from 2003-2012 and plot the medians of the annual estimated $\alpha$'s of these three groups for the subsequent  10-year windows (2003-2012, $\cdots$, 2010-2019). In this example, the performance of the skilled funds and the unskilled funds are separated during the subsequent 10-year windows. For the non-selected funds, it may contain some skilled funds as well as some unskilled funds. Nevertheless, our selected skilled funds still dominate the performance of the non-selected group over time, and the non-selected group further dominates the unskilled group. However, such distinction among the three groups narrows over time, indicating that the predictive power of our procedure deteriorates over a long horizon. 
More results are in Supplementary Material~H. Generally,  the medians of the estimated annual $\alpha$'s for the unskilled group are all negative, while the skilled group generated positive medians for the majority of the scenarios. In some favorable cases, for the selection based on the data from 2001-2010 and 2003-2012, the median for the selected skilled funds is above $2.5\%$. Such a pattern remains through the other 10-year window until 2010-2019.

\begin{table}[!ht]
\caption{\footnotesize Average annual expense ratios and turnover ratios of funds in skilled, unskilled and non-selected groups.}\label{table:exp.turn}
\vspace{-0.6cm}
\begin{center}
\footnotesize 
\begin{tabular}{c|l|ccc}
\hline\hline
             &   & Skilled     & Unskilled     &Non-selected\\     
\hline      
\multirow{2}{*}{2000-2009} 
&Expense (\%) &1.37 & - & 1.41\\
&Turnover (\%) &76.2& - & 80.5 \\
\hline
\multirow{2}{*}{2001-2010} 
&Expense (\%) & 1.38& - &1.40 \\
&Turnover (\%) & 91.1 & - & 85.6\\
\hline
\multirow{2}{*}{2002-2011} 
&Expense (\%) &1.34 &   2.09 & 1.38 \\
&Turnover (\%) & 67.6 &  510.4  & 85.1 \\
\hline
\multirow{2}{*}{2003-2012} 
&Expense (\%)& 1.14    &      1.87    &         1.36\\
&Turnover (\%) & 74.5  &    513.5  &   79.4 \\
\hline
\multirow{2}{*}{2004-2013} 
&Expense (\%) &   0.95     &    1.61      & 1.34\\
&Turnover (\%) & 51.6   &       192.5    & 76.1 \\
\hline
\multirow{2}{*}{2005-2014} 
&Expense (\%) &      1.21 &      1.72     &         1.31\\
&Turnover (\%) &  83.4      &     237.7    &      77.3\\
\hline
\multirow{2}{*}{2006-2015}
&Expense (\%) & 1.27  & 1.95 &  1.29\\
&Turnover (\%) &  52.3 &   2659.8     & 77.5\\
\hline
\multirow{2}{*}{2007-2016}
&Expense (\%) &  1.14     &      1.73     &         1.24\\
&Turnover (\%) & 34.9       &      441.6     &      77.4\\
\hline
\multirow{2}{*}{2008-2017} 
&Expense (\%) & - &  1.49  &  11.5\\
&Turnover (\%) & - &  103.2 &  73.5\\
\hline 
\multirow{2}{*}{2009-2018}
&Expense (\%) & 2.00 &  1.53    & 1.16 \\
&Turnover (\%) & 65.6  &  116.1 & 74.0 \\
\hline\hline
\end{tabular}
\end{center}
\end{table}

We also analyze the difference of behaviors in these three groups (skilled, unskilled, non-seleted) through expense ratios and turnover ratios provided by CRSP database.  An expense ratio is the ratio of the total investment that fund managers pay for the fund's operating expense. For each 10-year window, we calculate the average annual expense ratio for funds in each group. Note that we analyze fund $\alpha$'s net of trading fees, costs, and expenses, so the performance of funds should be largely affected by fund expenses. \citet{Carhart97} documents that equity funds generally underperform the benchmarks after expenses, which indicates that funds with low expense ratios are more favorable. Therefore, we expect skilled groups to have relatively lower expense ratios, and unskilled groups to have relatively higher expense ratios. The results in Table~\ref{table:exp.turn} agrees with our expectation for most of the time. This exhibits the influence of expense on the performance of mutual funds. 

On the other hand, we are interested in how activeness affects fund performance. We calculate the average annual turnover ratio to measure the activeness for each group. In the CRSP database, the turnover ratio for a fund is computed as the minimum of aggregated sales or purchases of securities over the average monthly total net assets of the fund. \citet{Carhart97} demonstrates turnover's negative effect on funds' performance. In Table~\ref{table:exp.turn}, the distinction in turnover ratios between the skilled and the non-selected groups is slight, but the unskilled group has much higher turnover ratios for all the 10-year windows, which is consistent with the argument in \citet{Carhart97}. Though high turnover ratios are not equivalent to bad performance, they could generate larger trading costs, fees, and expenses, which also explains higher average annual expense ratios for the unskilled group. It is worth pointing out that the skilled funds selected by the procedure in \citet{BSW10} have similar turnover ratios as the unskilled funds, which indicates that their skilled group has different behaviors compared to our selected group. This difference is caused by the distinct rankings of funds based on d-values and p-values, which will be discussed in Section~\ref{sec.sub:comp.p.p}. By and large, our results suggest that overactive mutual funds during a certain period should be cautiously treated or even avoided by investors.

\subsection{Comparison of d-value and p-value }\label{sec.sub:comp.p.p}
The selection of funds based on our d-values is quite different from those based on the p-value, e.g., BH procedure or Storey procedure. The p-values of our selected skilled funds are not necessarily small, indicating that the funds located on the right tail part of the empirical distribution of $\bZ$ are not necessarily skilled. To present the distinction, for a certain 10-year window, we collect the funds with the top 50 smallest p-values (denoted as p-group) and the funds with the top 50 smallest d-values (denoted as d-group) after excluding the intersection part. We compare these two groups by p-values, d-values, estimated annual $\alpha$'s, and future returns in the following discussion.

\begin{figure}[ht!!!]
\caption{\footnotesize The comparison of the p-group and d-group  selected based on the data from 2000-2009: (a) the box plots of p-values and d-values,  and (b)  the estimated annual $\alpha$'s for each 10-year window after that period. (``others" refers to  funds without top 50 smallest d-values nor p-values.)}
\centering
 \begin{subfigure}[b]{0.43\textwidth}
    \centering
    \includegraphics[width=\textwidth]{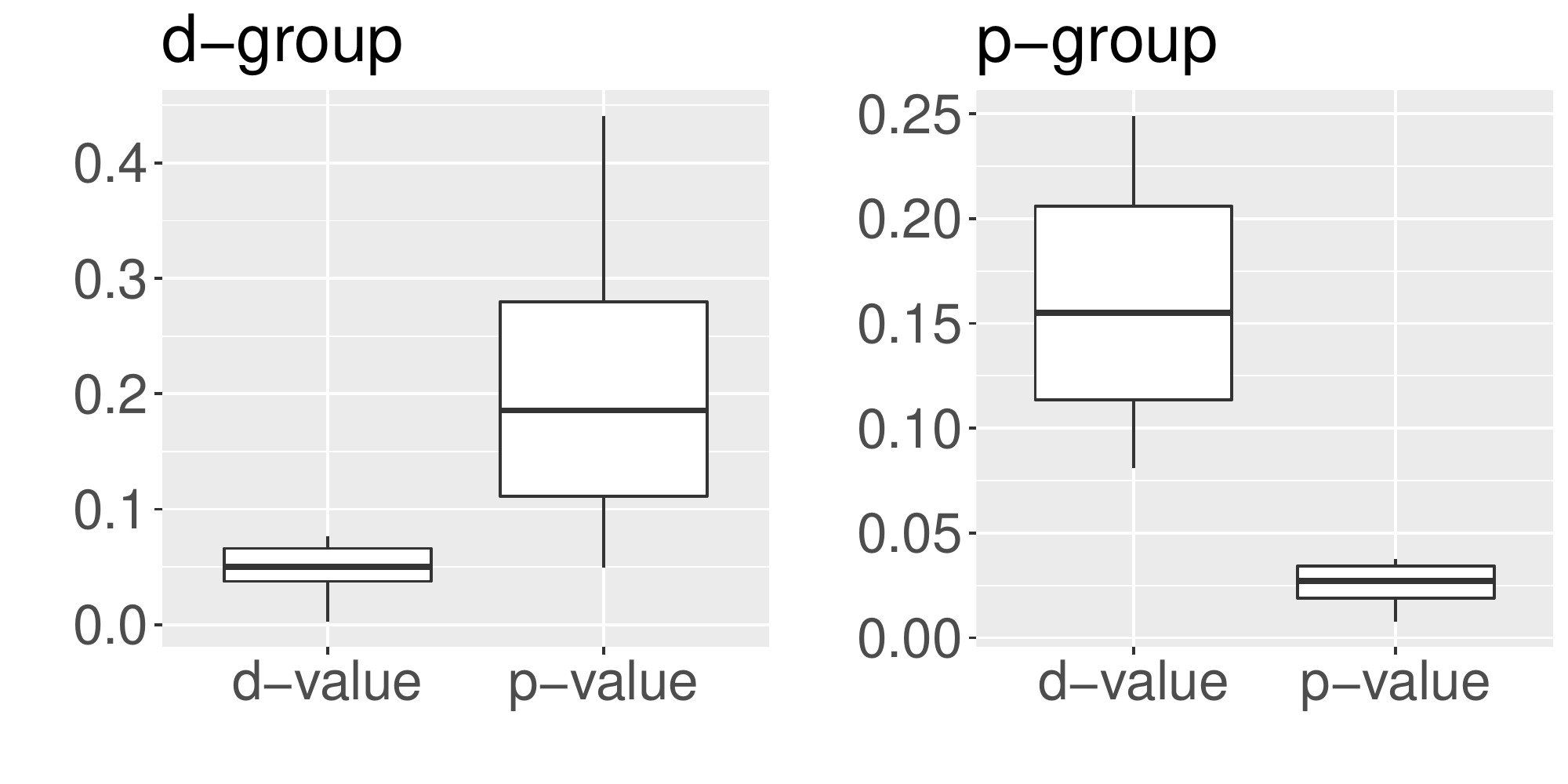}
    \vspace{-0.4cm}
    \caption{}\label{00_09_pp_1}
\end{subfigure}
\hspace{1.5cm}
\begin{subfigure}[b]{0.43\textwidth}
    \centering
    \includegraphics[width=\textwidth]{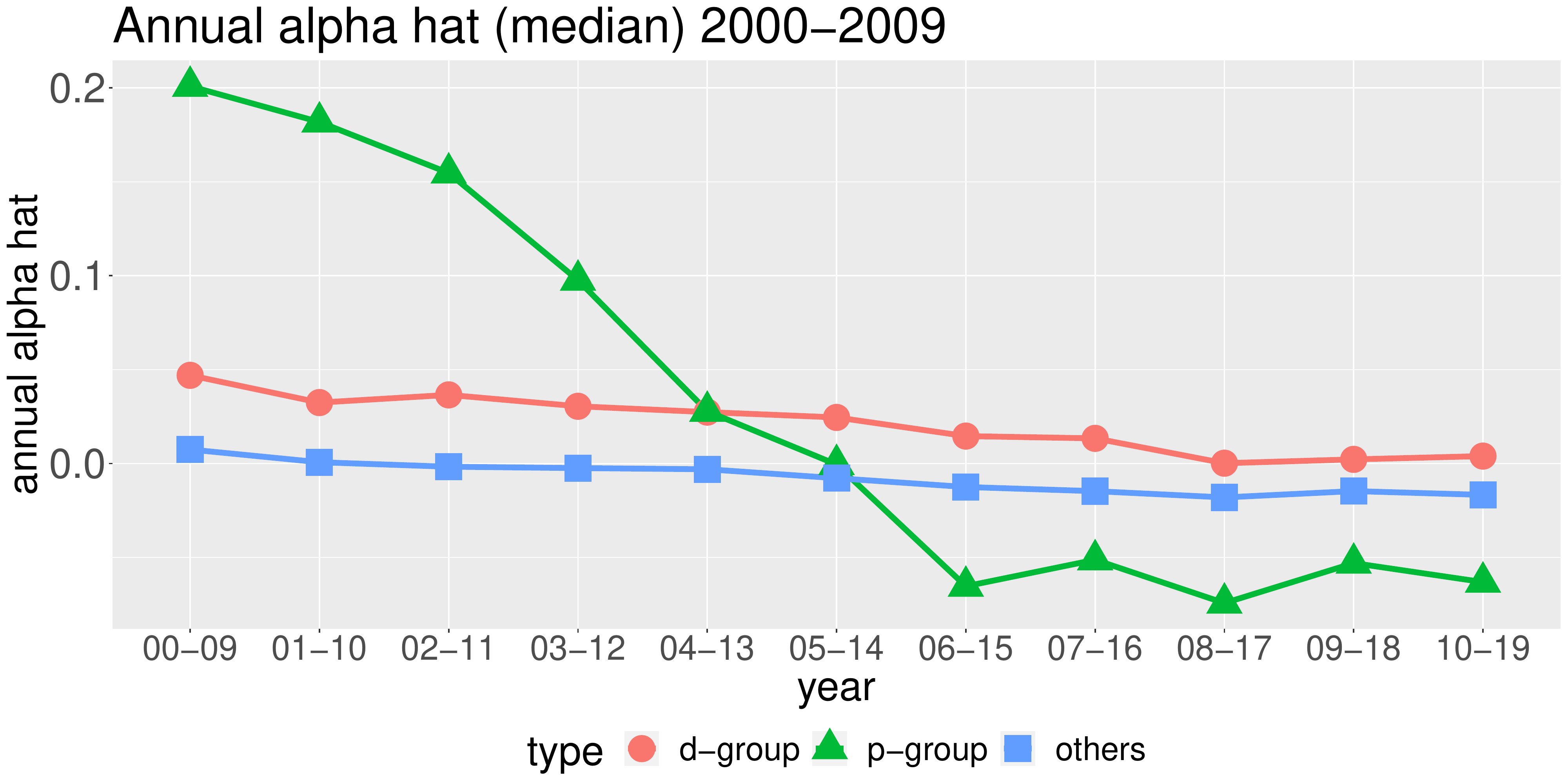}
    \vspace{-0.4cm}
    \caption{}\label{fig:00_09_differ_1}
\end{subfigure}
\end{figure}
 
Figure \ref{00_09_pp_1} provides the example boxplots of p-values and d-values for the two groups based on the data from 2000-2009. The d-group has a median p-value equal to 0.19, and even the minimum p-value is greater than 0.04. In contrast, all the p-values in the p-group are less than 0.04. Similar phenomenons can be observed in the boxplots for other periods in Figure~S5 (Supplementary Materials~I). Therefore, the rankings for funds based on d-values and p-values have a sharp distinction. A natural question is which ranking is better.  We answer this question by comparing the fund performance in the p-group and d-group in Figure~\ref{fig:00_09_differ_1}. The plot of the medians of estimated annual $\alpha$'s shows that the superior performance of the d-group is persistent in the subsequent moving 10-year windows compared to that of the p-group. For the p-group, we observe excellent performance in the beginning, but the performance deteriorates and becomes even worse than that of funds with relatively large p-values. More examples are shown in  Figure~S6 (Supplementary Materials~I). 
 
\begin{figure}[!ht]
\caption{\footnotesize The boxplots for cumulative returns of funds in p-groups and d-groups for the following years. D-values and p-values are computed based on the data during (a) 2000-2009 and (b) 2003-2012.}\label{00_09_return}
\centering
\begin{subfigure}[b]{0.32\textwidth}
    \centering
    \includegraphics[width=\textwidth]{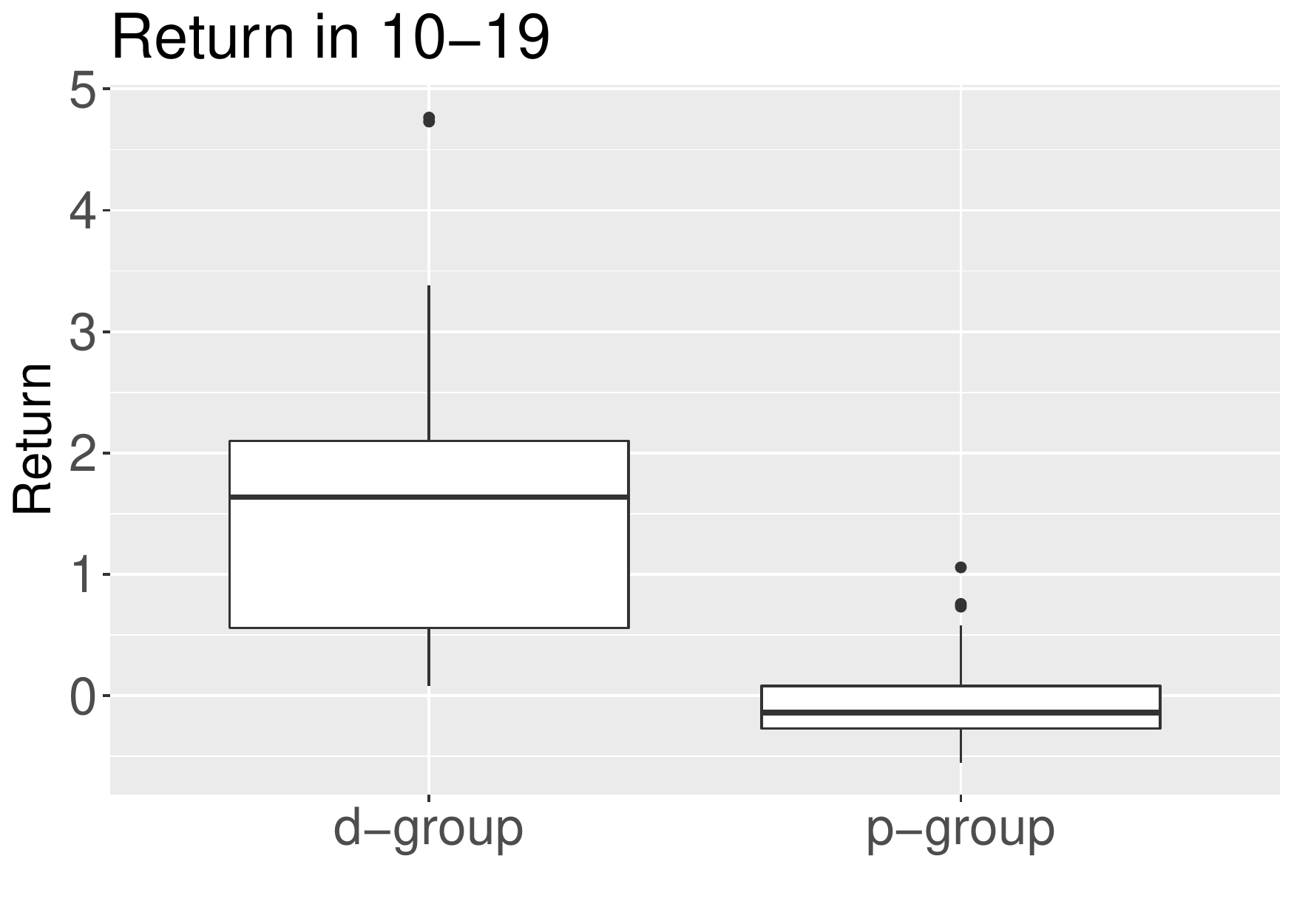}
    \vspace{-0.8cm}
    \caption{\footnotesize 2000-2009}\label{00_09_return_1}
\end{subfigure}
\hspace{2cm}
\begin{subfigure}[b]{0.32\textwidth}
    \centering
    \includegraphics[width=\textwidth]{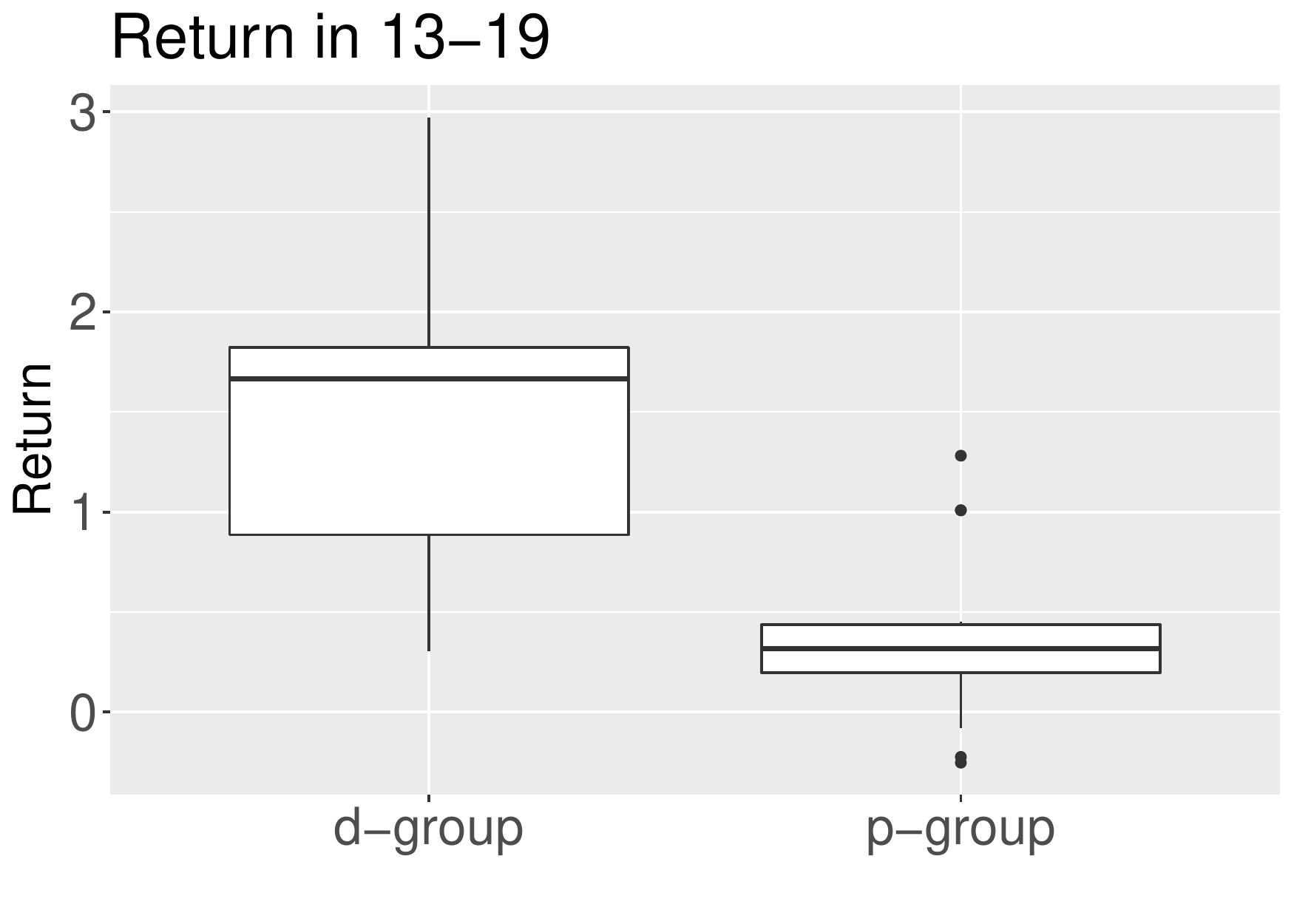}
    \vspace{-0.8cm}
    \caption{\footnotesize 2003-2012}\label{00_09_return_2}
\end{subfigure}
\end{figure} 

We also look at the cumulative returns of funds in the two groups for the subsequent years over a long horizon. Figure~\ref{00_09_return} contains example boxplots for the cumulative returns of funds in p-groups and d-groups. In Figure~\ref{00_09_return_1}, the cumulative returns of funds in the d-group, constructed based on the data from 2000-2009, have a median greater than 150\% during 2010-2019, but the cumulative returns of the p-group have a negative median.  More results are shown in Figure~S7 (Supplementary Materials~I). Overall, d-groups strongly outperform p-groups. Further investigation of the selected groups reveals additional support for our procedure. For instance, our selection of the d-groups based on 2003-2012 includes some prominent funds such as \verb!Fidelity! and \verb!ProFunds!, which has outstanding long-term performance in addition to their short-term performance for the year 2013. For example, \verb!Fidelity Select Portfolios:!\verb!Medical Equipment! \& \verb!Systems Portfolio! generated a total return of $297\%$ from 2013 to 2019 while \verb!ProFunds:!\verb!Biotechnology UltraSector! generated a total return of $236\%$ during the same period. However, such funds are not in the p-group constructed based on the data from 2003-2012. 

\section{Further Discussion}\label{sec:discussion}

The current paper focuses on the performance of the mutual funds net of trading costs, fees, and expenses. We may look at returns of mutual funds before expenses. The latter directly reflects mutual fund managers' capabilities to pick stocks. We can compute expenses in terms of the expense ratios, and add the estimated expenses to the current net returns. From the investors' side, despite the management fee that investors have to pay out, they also receive dividends from the investment of some funds in addition to the returns. Some dividend-oriented funds may pay a substantial amount of dividends, which can not be ignored for calculating profits. In our aforementioned analysis, we did not collect the information of dividends for each fund. The above factors motivate us to collect the related information for the mutual fund data in our future research. 


\section*{Acknowledgement}
We want to thank the Joint Editor, the Associate Editor and the reviewer for many constructive comments which substantially improve the quality of the paper. 

\spacingset{1}

\end{document}